\documentclass[%
 reprint,
 amsmath,amssymb,
 aps,
]{revtex4-2}
\usepackage[english]{babel}
\usepackage{dcolumn}% Align table columns on decimal point
\usepackage[utf8]{inputenc}
\usepackage[pdftex, pdftitle={Article}, pdfauthor={Author}]{hyperref} % For hyperlinks in the PDF
\usepackage{graphicx}
\usepackage{bm}% bold math
\usepackage{array,multirow}
\usepackage{amsmath}
\usepackage{amssymb}
\usepackage{color}
\usepackage{tabularx}
\usepackage{hhline}
\usepackage{multirow}
\usepackage{longtable}
\usepackage{booktabs}
\DeclareUnicodeCharacter{2212}{-}

\begin{document}
\preprint{APS/123-QED}
%\begin{frontmatter}

\title{Systematic study of amorphous 2D graphene, silicene and silicon
    carbide: Investigation of structural, electronic, optical and
    vibrational properties} 

%\thanks[footnoteinfo]{This paper was not presented at any IFAC meeting. Corresponding author M.~T.~Cicero. Tel. +XXXIX-VI-mmmxxi. Fax +XXXIX-VI-mmmxxv.}

%\author[]{E G\"urb\"uz$^1$, B Sanyal $^1$}
%\address{$^1$ Division of Materials Theory, Department of Physics and Astronomy, Box 516, Uppsala University, SE-751 20 Uppsala, Sweden}

%\ead{emel.gurbuz@physics.uu.se; biplab.sanyal@physics.uu.se}
\author{E. G{\"u}rb{\"u}z}
   \email[Correspondence email address: ]{emel.gurbuz@physics.uu.se}
    \affiliation{Division of Materials Theory, Department of Physics and Astronomy, Box 516, Uppsala University, SE-751 20 Uppsala, Sweden}
\author{B. Sanyal}
    \email[Correspondence email address: ]{biplab.sanyal@physics.uu.se}% Your name
    \affiliation{Division of Materials Theory, Department of Physics and Astronomy, Box 516, Uppsala University, SE-751 20 Uppsala, Sweden}

\date{\today}% It is always \today, today,

%\begin{keyword}                           
%\end{keyword}                             

\begin{abstract}  
Although two dimensional (2D) solids in their crystalline form and their van der Waals (vdW) heterostructures have generated a lot of attention in recent years, the exploration of their amorphous forms has not been done to a great extent. Here, we present a detailed analysis of structural, electronic and thermal properties of 2D amorphous graphene(A-Gra), silicene(A-Si) and silicon carbide(A-SiC) using Classical Molecular Dynamics (MD) simulations for structure generation, stability tests, thermal conductivity and vibrational analysis  while we use first-principles density functional theory based calculations for the calculations of electronic structure and charge distributions. It is found that A-Gra is planar at 0 K while it gets wrinkled at 300 K similar to its crystalline counterpart. At both temperatures, A-Gra can form stable vdW solids. A-Gra is found to be metallic with a thermal conductivity around 55.30 W/K.m. We also observe that A-Gra can absorb light from IR to UV range and has plasmon peaks in UV range but red-shifted compared to the crystalline graphene.  A-Si prefers to create covalent bonds in layered structures while it keeps its metallicity. A-Si's thermal conductivity is calculated as 2.68 W/Km which is the same value for amorphous one dimensional Si nanowire. A-Si's optical absorption range is found from IR to UV, but red-shifted relative to crystalline silicene's perpendicular components, whereas blue-shifted in parallel polarization relative to crystalline silicene. Its plasmon peak is found in UV range with 6 times higher intensity compared to crystalline silicene. We also demonstrate that two dimensional silicon carbide can be stabilized not only as a single layer but also in layered structures with bonding between Si atoms only. A-SiC is metallic and its thermal conductivity is found to be 70.29 W/K.m, higher than A-Gra. Its absorption range is from IR to UV with a red-shifted spectrum compared to its crystalline counterpart and plasmon peak residing in UV range. Another important finding is that except A-Gra, A-Si and A-SiC has no local vibrational modes, and all three structures' common heat carriers are extendons, especially diffusions.  Finally, the uneven charge distributions around the local ring structures in all three systems can be exploited in future electronic, thermo-electric, opto-electronic devices with their local functionalility with the advantage of their metallicity and low thermal conductivity.
\end{abstract}

\maketitle

\section{Introduction}
The usage of amorphous materials in electronics is  increasing day by day due to their superior electronic, mechanical and thermal properties compared to their crystalline counterparts. One of the most significant aspects of amorphous materials is their low cost and large-scale producibility to be used in a variety of different applications such as electronic devices, thin film transistors, catalysis, magnetic recording applications, energy storage etc. \cite{Am-general,Zachariasen,all} Low dimensional materials are also rising stars of the electronic industry by functionalization of 2D materials or creating combinations of these low dimensional structures, since the first discovery of graphene in 2004. \cite{2DA-Ex-GraTrend,2DCrystaloverview,2DCrystaloverview2} The increasing significant demand promoted research for improvement of designing semiconducting and semimetallic  low dimensional and van der Waals solids  as device components. \cite{2DvdW,2DvdW1} Starting from the first description of 2D glass made by Zachariasen, there are a great number of methods to produce continuous and fully random networks that can combine the unique properties of both amorphous and low dimensional materials. \cite{Zachariasen,Zachariasen-2,2DA-Exp-IOP,2DA-Ex-JAP,2DA-Ex-Am-Chalcogen} Since the last decade, many inorganic stable amorphous 2D materials with an atomic layer thickness; such as amorphous-C; amorphous-BP, amorphous-TMD, amorphous-BN, amorphous-MoS$_{2}$, etc., are being synthesized  successfully with different techniques such as exfoliation, electron irradiation, chemical vapor deposition (CVD) or physical vapor deposition (PVD). \cite{2DA-Ex-SciRep,AGra-Nature,2DA-Ex-Am-MoS2,2DA-Ex-Am-B.phosphor,2DA-Ex-Am-AgraCathod,2DA-Ex-Am-AgraOil,2DA-Ex-Am-MoS2-hydrogen,2DA-Ex-Am-MoS3-toxic,2DA-Ex-Am-Chalcogen,2DA-Ex-Am-BN-Nano}

Furthermore, prediction of possible low dimensional amorphous structures such as amorphous-graphene\cite{2DM-Agra,Nano-Agra,AGra-Nature,Carbonmech-Agra,2DM-2El,2d-2019-Agra,CGlass-struct}, amorphous-silicene\cite{Si1,Si2,Si3,Si-penta,Si-structure,Si-structure2}, amorphous-silicon carbide\cite{SiC-thermal,SiC-nanoribbon1,SiC-nanoribbon2}, amorphous-germanene\cite{2DA-Germanene}, amorphous-BN\cite{2DA-Ex-Am-BN}, etc. has become possible with molecular dynamics simulations to get a deeper insight of the amorphous structure's effect on mechanical, thermal and electronic properties. Mortazavi et al. showed the stiffness of the thermally low conducting amorphous graphene.\cite{Carbonmech-Agra} Zhu et al., studied the effects of disorder on the localization of the vibrational modes of amorphous graphene by comparing to Stone-Wales defect density in crystalline graphene, to understand deeper the low transport mechanisms of both graphene forms which can be used for thermal coatings and thermoelectric materials and many other applications.\cite{Nano-Agra} Amorphous graphene's low thermal conductivity was also studied by quantifying the amorphousness in comparison to the sp$^2$ hybridization in crystalline graphene \cite{2DM-Agra} whereas the effects of sp$^3$ to sp$^2$ hybridizations on thermal conductivity of amorphous carbon in 3D were studied studied previously.\cite{3D-Mech-Agra} Moreover, Zhou et al., showed the structure and dimension dependent amorphous silicon's unusually higher thermal conductivity then nano-grains including crystalline counterparts.\cite{Si2,Si3}     

Although a few 2D amorphous structures have been synthesized, their structures and properties haven't been understood as much as 2D-crystalline materials. In this article, we focused on a systematic study of amorphous 2D- graphene (A-Gra); silicene (A-Si) and silicon carbide (A-SiC). We made a detailed structural analysis that can be a benchmark between classical molecular dynamics and first principles calculation to use a reliable system size. Following the structural analysis, we calculated electronic and optical properties of amorphous monolayers. Thermal conductivity and the analysis of dominant heat carriers were done through the calculations of vibrational modes. Moreover, we studied disordered layered systems and investigated vdW solids.  Our findings show that the studied structures; A-Gra, A-Si and A-SiC's metallicities, optical absorption from IR to UV range, plasmon frequencies located in UV  range  along with their low thermal conductivity make them good candidates to be used in future thermoelecric and thermal coatings, electronic, optoelectronic and in many more application areas.

\section{Methods}
The amorphous structures  were prepared by classical molecular dynamics (CMD) simulations with the open source LAMMPS package \cite{lammps} employing the Tersoff potential.\cite{tersoff} 
The different-sized structures were produced for all A-Gra, A-Si and A-SiC structures in the following supercells; 10$\times$10 containing 200 atoms and 50$\times$50 containing 5000 atoms  where the lattice parameter of the 2 atoms unit cell was considered as 2.46 \AA, 3.87 \AA and 3.10 \AA respectively. The vacuum region was taken to be 40 \AA.
The generation of all the structures was initiated by distributing the atoms randomly followed by heating at 5000 K; fast-quenching to 2000 K with a $51.28\times10^{14}$ K/s cooling rate; applying short equilibration at 2000 K for 135 fs; quenching to 300 K with a $49.71\times10^{13}$ K/s cooling rate and the final equilibrating step at 300 K for 3420 fs. We used \textit{Berendsen thermostat} \cite{berendsen} for temperature re-scaling with 45 fs relaxation time with 1 fs timestep.

The stability of the structures was tested by first applying force minimization at T = 0 K with a stopping energy tolerance of $1\times10^{-18}$ eV and secondly, applying Berendsen thermostat at 300 K for 1.6 ns with varying damping parameters from strong to weak(1 fs, 5 fs, 10 fs, and 1 ps). The structures' final lattice vectors obtained for the 10$\times$10 supercells are 24.52 \AA, 37.85 \AA and 30.40 \AA for A-Gra, A-Si and A-SiC respectively. For the 50$\times$50 supercell, the values obtained are 122.72 \AA, 190.25 \AA and 151.55 \AA. The corresponding number densities are $ 3.32 \times 10^{15}$ cm$^{-2}$, $ 1.38 \times 10^{15}$ cm$^{-2}$ and $ 2.18 \times 10^{15}$ cm$^{-2}$.

Following the production step, 10$\times$10 structures are used for electronic structure analyses by employing first-principles calculations using VASP (Vienna ab initio simulation package) \cite{vasp2,vasp3}. Projector-augmented-wave (PAW)\cite{blochl} formalism was employed for ion-electron interaction and the electron exchange and correlation (XC) potential was described by the Perdew-Burke-Ernzerhof (PBE) form within the generalized gradient approximation (GGA)\cite{pbe}. For the Brillouin Zone integration, we used a  $(1\times1\times1)$ set of k-points generated by the Monkhorst-Pack scheme\cite{monk}. The kinetic energy cutoff for the plane-wave basis set of A-Gra, A-Si, and A-SiC was set respectively as 520 eV, 320 eV and 840 eV. The total energies were minimized with an energy tolerance of $10^{-4}$ eV. All atomic positions and lattice constants were optimized by the conjugate gradient method. For bilayers and trilayers, the interlayer binding energy per atom (BE$_{interlayer}$) was calculated using Eq. \ref{eqn:Eq.1}:
\begin{equation}
\label{eqn:Eq.1}
BE_{inter-layer} = \frac{(N_{layer}\times E_{SL})-E_{ML}}{N_{atom} \times N_{layer}}  
\end{equation}
where $E_{SL}$ refers to the energy of the single-layer and $E_{ML}$ refers to the energy of the bi-/tri- layers. $N_{layer}$ is the number of constituent layers in the multilayer structure and $N_{atom}$ is the number of atoms per layer \cite{bindingenergy}.
\\
Optical properties of 10$\times$10 structures were calculated with independent particle approximation (IPA) only considering the interband transitions where exchange and correlation were not considered and the local field effects (LFE) coming from density variation of the Hartree potential were neglected \cite{optic1} because of our limited computational resources. The imaginary part of the frequency-dependent dielectric function ($\epsilon^{ij}_{2}$) at the long-wavelength limit was calculated with Eq. \ref{eqn:Eq.2} and the real part ($\epsilon^{ij}_{1}$) was calculated through Kramers-Kronig transformation with Eq. \ref{eqn:Eq.3}, where both the perpendicular(in-plane polarization which was represented as $ij = xx$)  and parallel(out of plane polarization which was represented as $ij = zz$) components were calculated by considering $\Gamma$ point ($1\times1\times1$) in the Brillouin zone. For calculating dielectric functions of crystalline graphene, silicene and silicon carbide, a 45$\times$45$\times$1 set of k-points generated by the Monkhorst-Pack scheme was used. The kinetic energy cutoff for the plane-wave basis set of graphene, silicene, and silicon carbide was set respectively to 520 eV, 320 eV and 520 eV. The total energies were minimized with an energy tolerance of $10^{-4}$ eV. 
\begin{widetext}
\begin{equation}
\label{eqn:Eq.2}
\epsilon^{ij}_{2} = \frac{4 \pi e^{2}}{\Omega} \lim_{q \to 0} \frac{1}{q^{2}} \sum_{v,c,\textbf{k}} 2w_{\textbf{k}} \delta (\epsilon_{c\textbf{k}}-\epsilon_{v\textbf{k}}-\omega) \times \langle  u_{c\textbf{k}+q\textbf{e}_{i}} \mid u_{v\textbf{k}} \rangle \langle u_{v\textbf{k}} \mid u_{c\textbf{k}+q\textbf{e}_{j}} \rangle
\end{equation}
\end{widetext}
where $\Omega$ represents the volume of the cell, $\textbf{e}_{i,j}$ are the unit vectors along three directions, $c$ and $v$ refer to conduction and valence states respectively, $\epsilon_{c\textbf{k}}$ is the conduction band energy and $\epsilon_{v\textbf{k}}$ is the valence band energy, $u_{c\textbf{k}}$ represents cell periodic part of the orbitals at the k-point \textbf{k}.
\begin{equation}
\label{eqn:Eq.3}
\epsilon^{ij}_{1} =  1 + \frac{2}{\pi} \int_{0}^{\infty}\frac{\epsilon^{ij}_{2}(\omega^{'})\omega^{'}}{\omega^{'^2}-\omega^{2}}d\omega^{'}
\end{equation}
The complex dielectric function is related to the refractive index($n$) and extinction coefficient ($\kappa$) as $n(\omega)+i\kappa(\omega)=\sqrt{\epsilon(\omega)}$. So extinction coefficient can be written as $\kappa(\omega)= ( \frac{\sqrt{\epsilon_{1}(\omega)^{2}+\epsilon_{2}(\omega)^{2}}-\epsilon_{1}(\omega)}{2})^{(\frac{1}{2})}$. Absorption coefficient has a relation to extinction coefficient as $\alpha(\omega)=2\omega\kappa(\omega)/c$, where $c$ is the light velocity in vacuum. \cite{Gra-SiC} Consequently refractive index is Eq. \ref{eqn:Eq.4},
\begin{equation}
\label{eqn:Eq.4}
n(\omega)= (\frac{\sqrt{\epsilon_{1}(\omega)^{2}+\epsilon_{2}(\omega)^{2}}+\epsilon_{1}(\omega)}{2})^{\frac{1}{2}}
\end{equation}
The electron energy loss spectrum (EELS) can be calculated in terms of the imaginary part of the dielectric function as Eq. \ref{eqn:Eq.5}
\begin{equation}
\label{eqn:Eq.5}
L(\omega)= -Im(\frac{1}{\epsilon(\omega)})=\frac{\epsilon_{2}(w)}{\epsilon_{1}(w)^{2}+\epsilon_{2}(\omega)^{2}}
\end{equation}
%\\
The lattice thermal conductivity of the 50$\times$50 structures was calculated based on Green-Kubo formula by applying equilibrium molecular dynamics shown in Eq. \ref{eqn:Eq.6}:
\begin{equation}
\label{eqn:Eq.6}
\kappa = \frac{V}{3k_{B}T^{2}} \ \int_{0}^{\infty} dt \Big\langle \textbf{J}(t) \cdot \textbf{J}(0)  \Big\rangle  \
\end{equation}
where $k_B$ is the Boltzmann constant, \textit{T} is the temperature and the angular brackets denote an autocorrelation function of heat flux \textbf{J}. The thermal conductivity of single-layer amorphous structures can be calculated as the average of x- and y- components because the sheet is isotropic along the x- and y- directions. Thus, the volume \textit{V} in Eq. \ref{eqn:Eq.6} is the system's volume, in the formula was calculated over surface area times nominal thickness which was kept constant as 3.35 \AA. For the sake of statistical accuracy, thermal conductivity was averaged over 6 independent simulations by assigning different initial atomic velocities. A-Gra's and A-Si's(in 4 ns) thermal conductivities were calculated with the following parameters;  time step 0.25 fs;  correlation length 100 fs and sample interval 50 fs whereas A-SiC's (8 ns) parameters. 
For a better understanding of low thermal conductivity, vibrational and mode participation ratio analysis were done for 50$\times$50 systems by using Eq. \ref{eqn:Eq.7} and Eq. \ref{eqn:Eq.8} employing the Fourier transformation of the velocity autocorrelation of every \textit{i}th atom over 250 ps with 1 fs time step.  

The vibrational density of states (VDOS) is calculated as
\begin{equation}
\label{eqn:Eq.7}
g(\omega)= \frac{1}{3NK_{b}T}\int_{-\infty}^{\infty} dt \sum_{1}^{N} \Big\langle \textbf{v}_{i}(t+\tau) \cdot \textbf{v}_{i}(\tau)  \Big\rangle e^{i\omega t}  
\end{equation}
The participation ratio (PR) is calculated as
\begin{equation}
\label{eqn:Eq.8}
PR(\omega)= \frac{1}{N}\frac{(\sum_{i}VDOS_{i}(\omega)^2)^2}{\sum_{i}VDOS_{i}(\omega)^4}. 
\end{equation}

\section{Results and Discussion}

\subsection{Structural Analysis}\label{sec21}

\begin{figure*}
 \centering
%\begin{figure}[h]\centering
 \includegraphics[width=\textwidth]{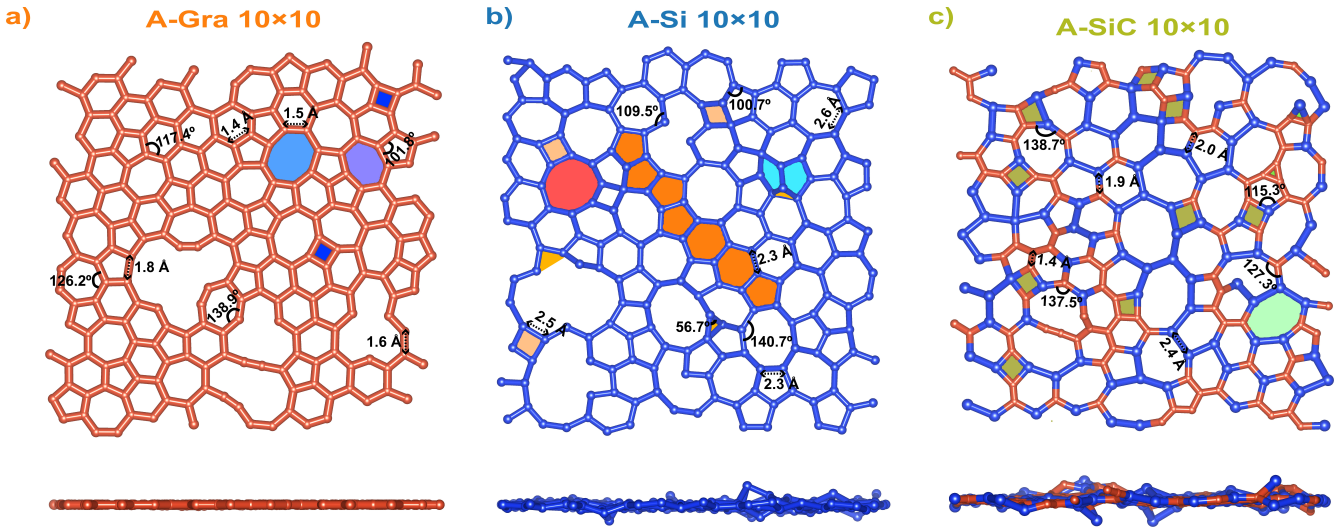}
\caption{The $10\times10$ structures, a) A-Gra, b) A-Si and c) A-SiC, side and top views are represented together with their varying bond lengths and bond angles at  0 K. A few of the system-specific rings are also represented with different color fills. A-Gra's the smallest 4-fold; and the larger 7-fold and 8-fold rings are marked. A-Si's varying ring size from 3-fold to 8-fold is marked and with orange color "chain-defect" is shown. A-SiC's existing 3-fold to 7-fold ring examples also show green shades.}
\label{fig1}
\end{figure*}

Structural analysis is the most commonly used technique to quantify the produced structures' amorphicity. The most used methods are radial distribution \textit{g(r)} and angular distribution functions. An amorphous mesh includes varying bond lengths, angles, and high porosity in contrast to a crystal mesh. The porous nature can be seen as rings in 2D amorphous counterparts. In Fig. \ref{fig1}, the 10$\times$10 A-Gra, A-Si and A-SiC amorphous structures' top and side views are shown along with the bond lengths, bond angles, rings and bonding types, after an ionic relaxation with density functional theory. Compared to the literature results of A-Gra \cite{2DM-Agra,Nano-Agra,2DM-2El,2d-2019-Agra,CGlass-struct,AGra-Nature} and A-Si \cite{Si1,Si-structure,Si-structure2,Si-penta}, our smaller cells show acceptable matches. Also, the dominant rings are different in A-Gra and A-Si, which is also in agreement with the literature. For example, A-Si shows a similar linear chain defect as seen in Fig. \ref{fig1},b. In this study, we also present the first analysis of 2D-A-SiC which is an interesting combination of A-Gra and A-Si at 0 K. The most distinguishing property between these three structures is that their planar or wrinkled nature seen from their side views. For example; A-Gra prefers to be planar in contrast to A-Si, while the most dominant non-planar nature is seen in A-SiC. 

\begin{figure*}
%\begin{center}
\centering
\includegraphics[width=\textwidth]{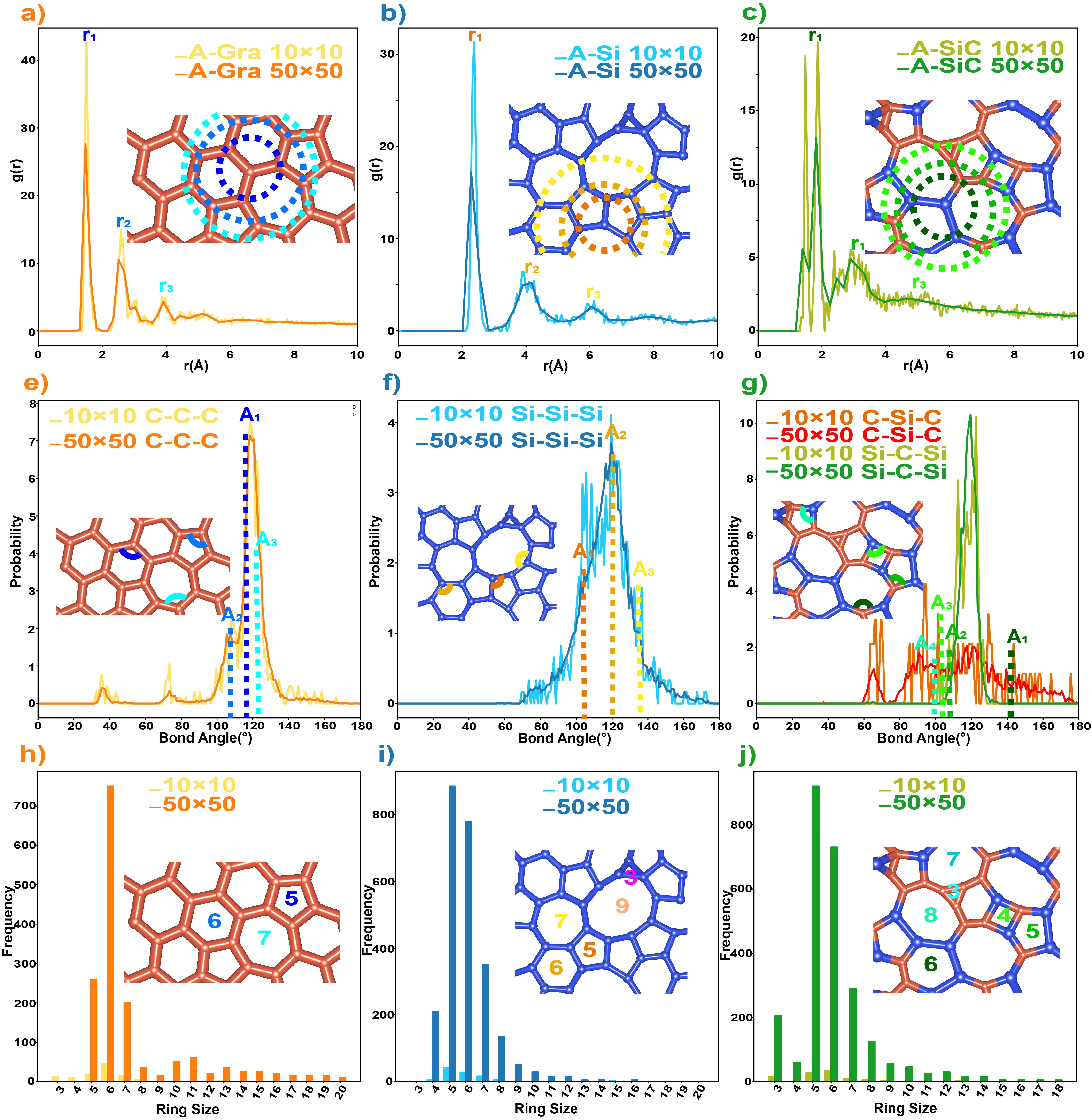}
\caption{By starting from a) to c), A-Gra; A-Si and A-SiC's rdf plots are represented for large(darker color) and small systems'(lighter color) sizes at 300 K.  From e) to g), the angular distribution of the systems in the same order is represented. From h) to j), ring analysis is represented in the same order. The insets are added from small systems at 0 K to show the agreement. The color codes are used to visualize the peaks on the insets.}
\label{fig2}
%\end{center}
\end{figure*}

The detailed analysis is presented in Fig. \ref{fig2}. Here we show the structural correspondence between the 50$\times$50 and 10$\times$10 structures at 300 K right after the production step with molecular dynamics simulations. 
In the first row of Fig. \ref{fig2}, we added a visual representation of each studied property at 0 K as insets to show the structural consistency between the two supercell size considered. Fig. \ref{fig2} \textbf{a}, \textbf{b} and \textbf{c} represent rdf plots of A-Gra, A-Si and  A-SiC respectively. Visibly, the noisy rdf peaks of small cells  are diminished by increasing the cell size, while the first, second and third peak positions are kept almost the same in both sizes. This proves the reliability and quality of the small cells. \textbf{r$_{1}$}, \textbf{r$_{2}$}, and \textbf{r$_{3}$} are shown as the radial distances between the atoms, which are from the randomly selected atom in the structure to the atoms sitting in the first, second and third neighborhood rings. An interesting observation in A-Gra's rdf as shown in Fig. \ref{fig2} \textbf{a}, is a slightly split peak of \textbf{r$_{2}$}, which has also been seen in previous studies. This may arise due to the complex polygonal network of C atoms. Interestingly, this feature is not observed in A-Si.
A-Si shows the same smoothing in the radial distribution function (rdf) denoted as g(r) with an expected shift towards the right in the rdf peaks compared to the A-Gra's rdf basically because of the atomic size difference. While the A-SiC shows a good combination of both C and Si atoms in the rdf with corresponding splittings in the first and second peaks. 
Fig. \ref{fig2} \textbf{d}; \textbf{e}; \textbf{f} represents the existing varying bond angles in the A-Gra; A-Si and A-SiC. A-Gra's varying bond angles agree with other studies that include the comparison to crystalline graphene. Carbon likes more hexagonal rings, we can see smaller, more distinct peaks compared to the A-Si. In Fig. \ref{fig2} \textbf{f}, we mostly focused on c-Si-C and Si-C-Si bonding angles' variation. The prominent property of C is kept in the Si-C-Si bonding case by showing more distinct narrower peaks compared to Si-centered C-Si-C which shows Si atoms property with wide single peak distribution.
Fig. \ref{fig2} \textbf{h}; \textbf{i}; \textbf{j} shows the favorite ring creation in A-Gra; A-Si; and A-SiC. The overall impression is that small cells can create small rings with 3 atoms compared to the bigger cells. The very big rings that are seen in the big cells can be sourced by the edges of cells by keeping the periodic boundary conditions conserved in the analysis and also big defects like rings can be observed during the production depending on the parameters. Carbon's favorite ring is undoubtedly hexagonal for graphene. The following favorite rings that exist in A-Gra  are hexagons, pentagons, and heptagons which are seen in crystalline allotropes of graphene. The A-Gra may show rings with 3 - 4 atoms or rings with 8 - 9 atoms, however, the dominant rings are respectively; hexagons; pentagons, and heptagons. A-Si's favorite rings can be formed as respectively pentagonal, hexagonal, heptagonal, and tetragonal. The rest rings are the minority in the overall structural analysis for A-Si. The dominant ring shape; the pentagon puts a conflicting structure compared to crystalline silicene.  This  may raise the question of the possibility that some pentagons included in silicene's  allotropes may exist with better stability and free-standing features and encourage further studies. Unlike A-Gra and A-Si, A-SiC rings analysis shows that 3 atoms including small rings can be favorite structures and can race with heptagons in the overall ring population in addition to the most dominant pentagon and the following hexagons. The heptagon dominance shows the silicon atoms are the main coordinator in the ring shape.

\subsection{Electronic Structures}\label{sec22}

\begin{figure*}
%\begin{center}
\centering
\includegraphics[width=\textwidth]{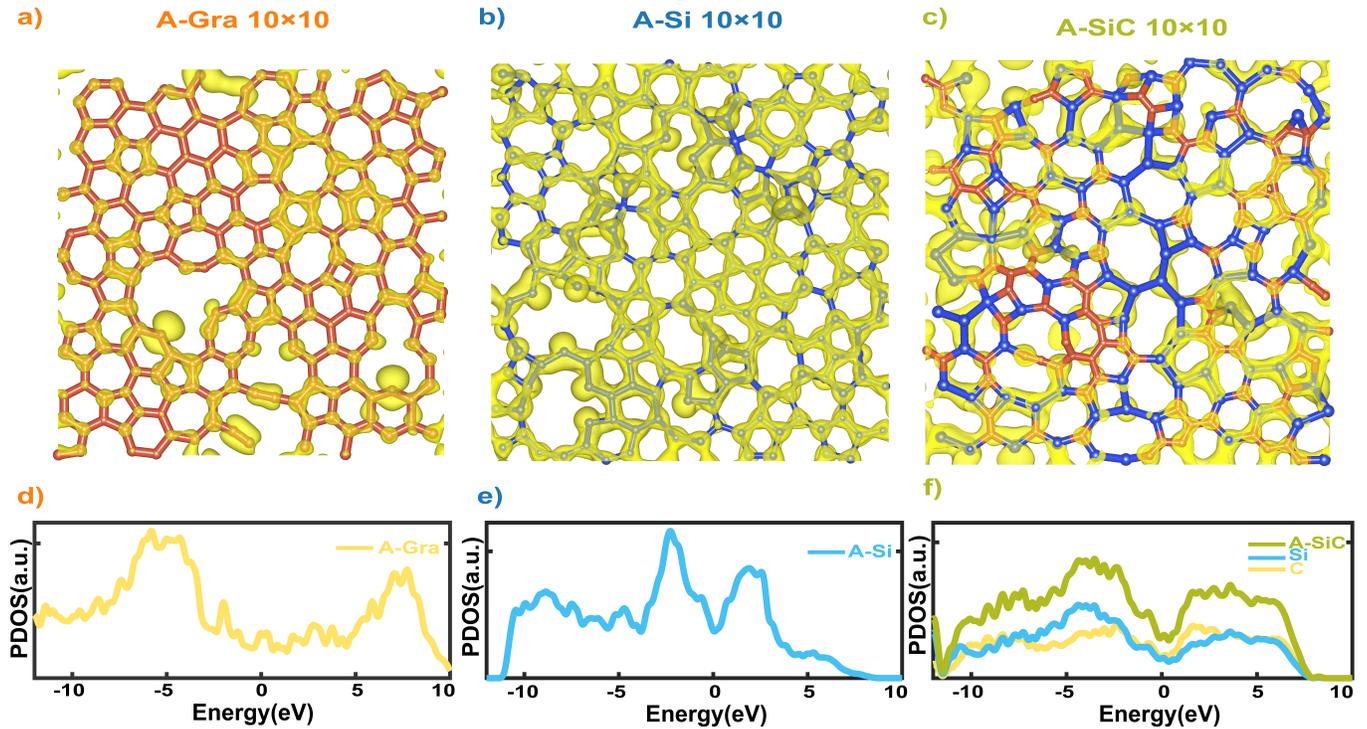}
\caption{Partial charge distributions at Fermi level of the small systems at  0 K is represented respectively a) A-Gra; b) A-Si; and c) A-SiC. The electronic density of states of the systems is represented in the same order from d to f. Atom projected density of states is also shown on f for A-SiC.}
\label{fig3}
%\end{center}
\end{figure*}

In this section, we present the electronic density of states and charge distributions of single layer A-Gra, A-Si and A-SiC as can be seen in Fig. \ref{fig3}. The general property of all the structures is that they are metallic. A-Gra's electronic structure has been studied and observed as metallic previously\cite{2DM-2El,AGra-Nature}. A previous study \cite{Si1} assumed A-Si as a semiconductor. On the contrary, we show that A-Si has a metallic nature. A-SiC's electronic nature is shown here for the first time. The charge distribution around fermi level shows quite different distributions from small ring sizes to big-defect-like ring sizes. In the case of A-Gra shown in Fig. \ref{fig3} a), the hexagonal rings have an equal charge distribution on atoms, while most of the pentagons, which are placed close to a big ring show increasing charge density. The big rings which have dangling bonds show a high charge density too. Furthermore, some atoms that show molecular binding (sp2 hybridization) with $~$ 180 $^{\circ}$ bonding angle in a big defect-like ring can show a big equally shared  charge density. We found that A-Gra and A-Si's charge density at around fermi energy arises from \textit{$p_{z}$} orbitals' contributions solely as in their crystalline semimetallic phases. Particularly, A-SiC's charge distribution around the fermi level has  a higher \textit{$p_{y}$} contribution and \textit{$p_{z}$} orbitals' contributions while crystalline SiC has a 2.55 eV gap \cite{SiC2023}. 
In contrast to A-Gra, A-Si shows a large charge density overall for all ring types. Some Si atoms on the edge of a large ring and having dangling bonds show a comparably larger charge on them, which is also larger than the A-Gra's big rings' dangling bonds' charge densities. Whereas in A-SiC, both Si and C atoms show different charge densities because of varying hybridization depending on the neighboring atoms. In the case of isolated rings, C or Si atoms can show a similar charge distribution in A-Gra or A-Si.   
Now we discuss the bilayer and trilayer amorphous structures. For a bilayer, we find that a 90 $^{\circ}$ rotated layer with respect to the other yields lower energy than the unrotated one. We tested these two cases for all structures as shown in Table \ref{table1}. Interestingly, A-Gra keeps the van der-Waals (vdW) stacking type in A-Gra case. The calculated layer binding energy per atom Table \ref{table1} agrees with the crystalline vdW bonded bilayer graphene\cite{bindingenergy,bindingenergy2}. In addition to the test case, we saw the same behavior at 300 K with force field calculations. However, A-Si prefers to create a covalent bonding between the layers. The calculated binding energy is found to be 18 times higher than A-Gra for the cases where the layer is placed identically on the top of the other layer, and 16 times higher than A-Gra for the rotated cases. Covalent bonding between the layer causes the increasing binding energy per layer. More interestingly, we observed that while A-SiC's Si atoms create covalent bonding with the upper layer's Si atoms, C atoms prefer to keep the vdW bonding with the upper layer's C atoms. Another possible Si-C bonding type rarely creates weak bonds according to the buckling position of atoms. The calculated binding energies of A-SiC layers per atom were calculated as smaller than A-Si but obviously larger than A-Gra layers, which is because of the intermediate bonding type. Next, we studied trilayer structures as shown in Table \ref{table1} and we found a striking similarity in bonding with the bilayers. To summarize this section, A-Gra layers can show planar vdW stacking at 0 K. At 300 K while the layers lose planar shape, they still keep vdW stacking. A-Si would like to create covalent bonding type between layers. Finally, A-SiC includes an interesting combination of covalent bonding type (Si-Si), weak bonding (Si-C), and vdW bonding (C-C). All the structures keep their metallicities.
%We calculated these  systems' energies as respectively -4415.62 eV for A-Gra, and -3751.49 eV for A-Si. 

\begin{table*}[ht]
\begin{center}
%\begin{ruledtabular}
\begin{tabular}{ | c | c | c | c | c | c | }
\hline

\multicolumn{3}{|c|}{} & A-Gra & A-Si & A-SiC \\
\hline
%Single-layer & -1767.06 & -950.69 & -1328.98  \\
%\hline
%Binding Energy(meV) & 0 & 0 & 0  \\
%\hline
%\hline
%Double-layer & -3542.57 & -2052.71 & -2727.82  \\
%\hline
\multirow{12}{*}{\rotatebox{90}{ BE$_{interlayer}$(meV)}}& \multirow{10}{*}{\rotatebox{90}{Double-layer}} & \multirow{4}{*}{Identical} & 21.13 & 378.35 & 174.68 \\
& & & &  \includegraphics[width=1.0in]{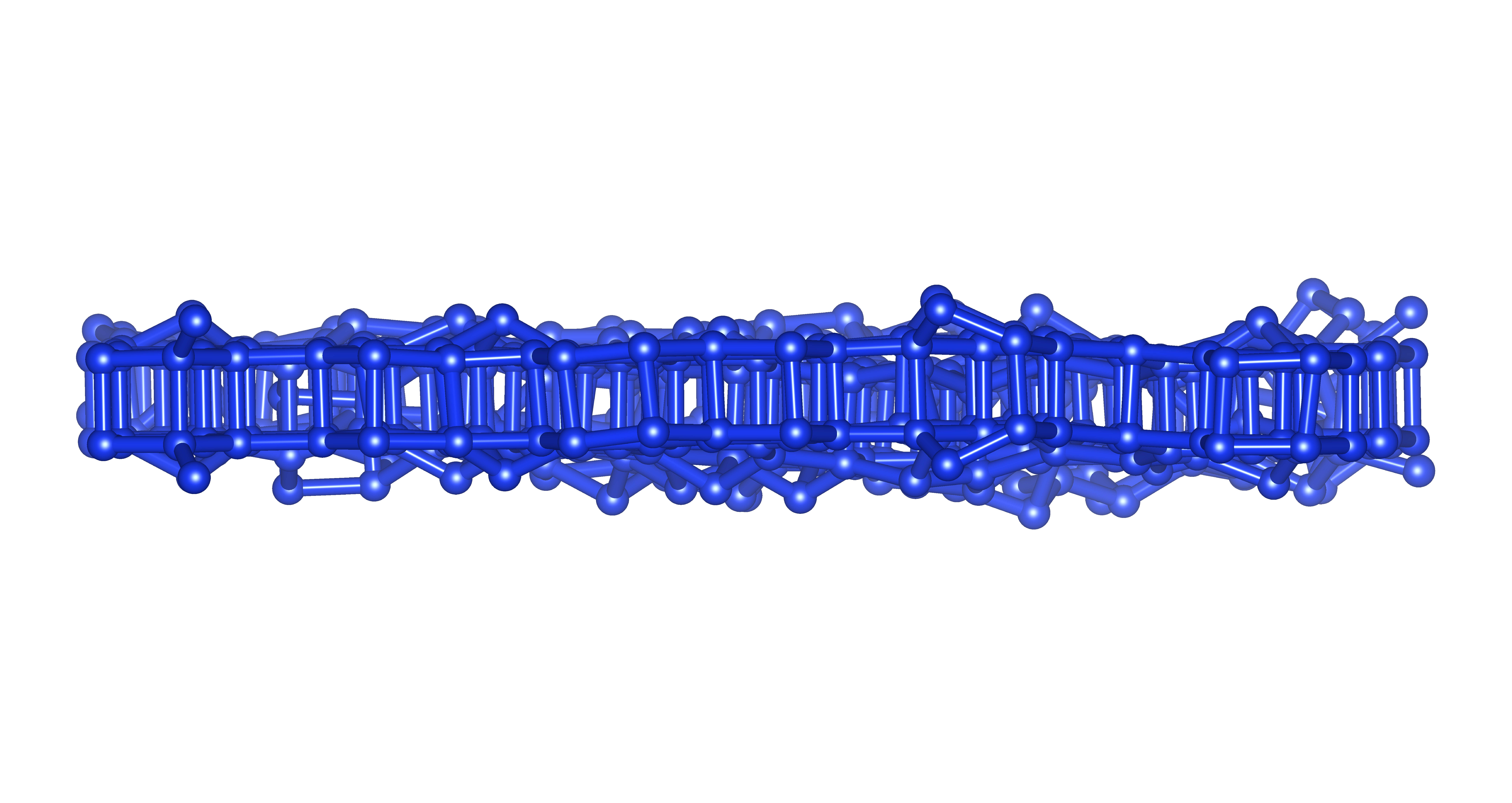} & \includegraphics[width=1.0in]{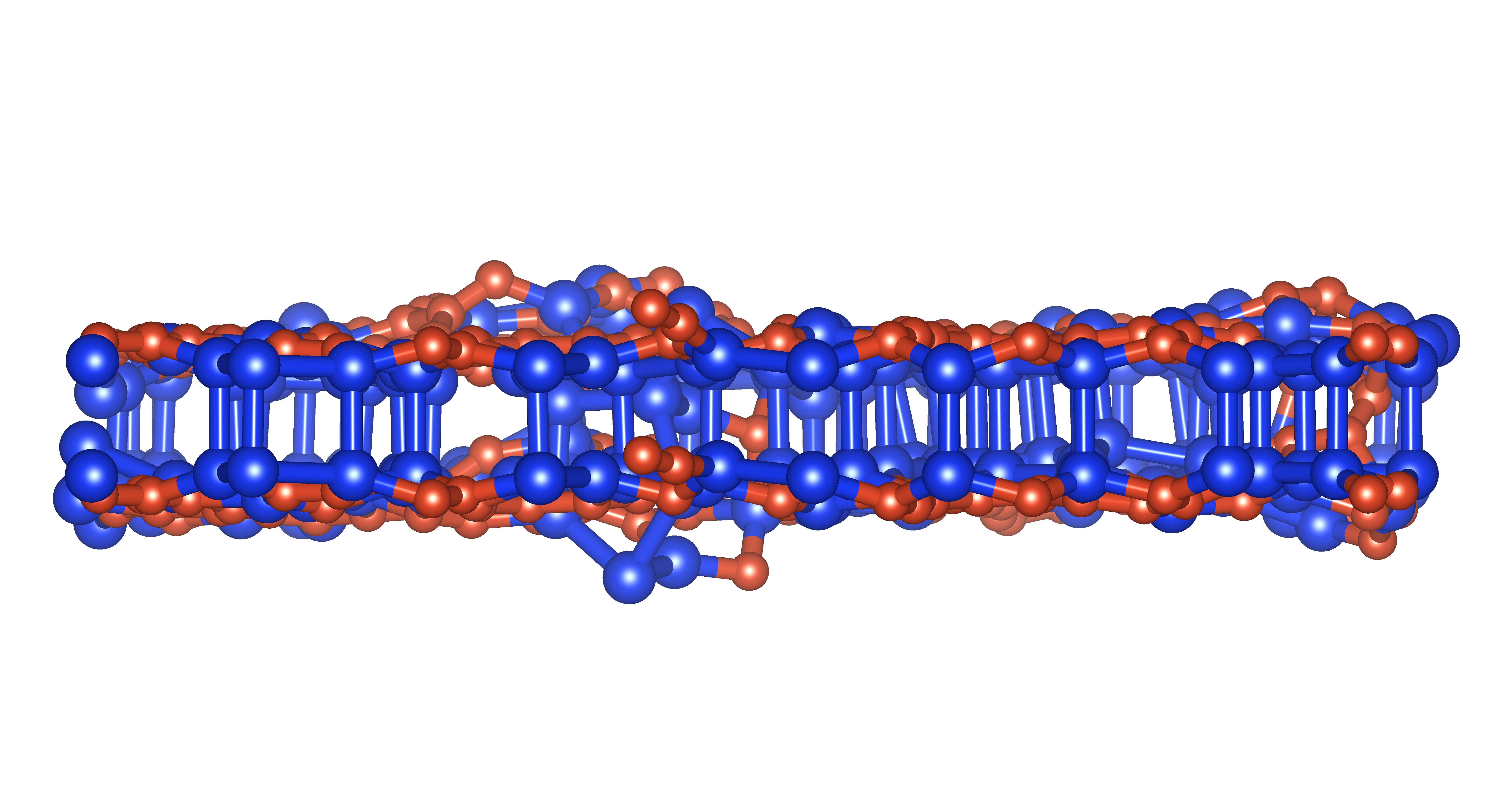} \\
\hhline{|~|~|-|-|-|-|}
%\cline{3-6}
%\hline
%\hline
%Rotated Double-layer & -3542.82 & -2042.90 & -2726.82  \\
%\hline
%\hline
& & \multirow{4}{*}{Rotated} & 21.75 & 353.83 & 172.18  \\
& & &  \includegraphics[width=1.0in]{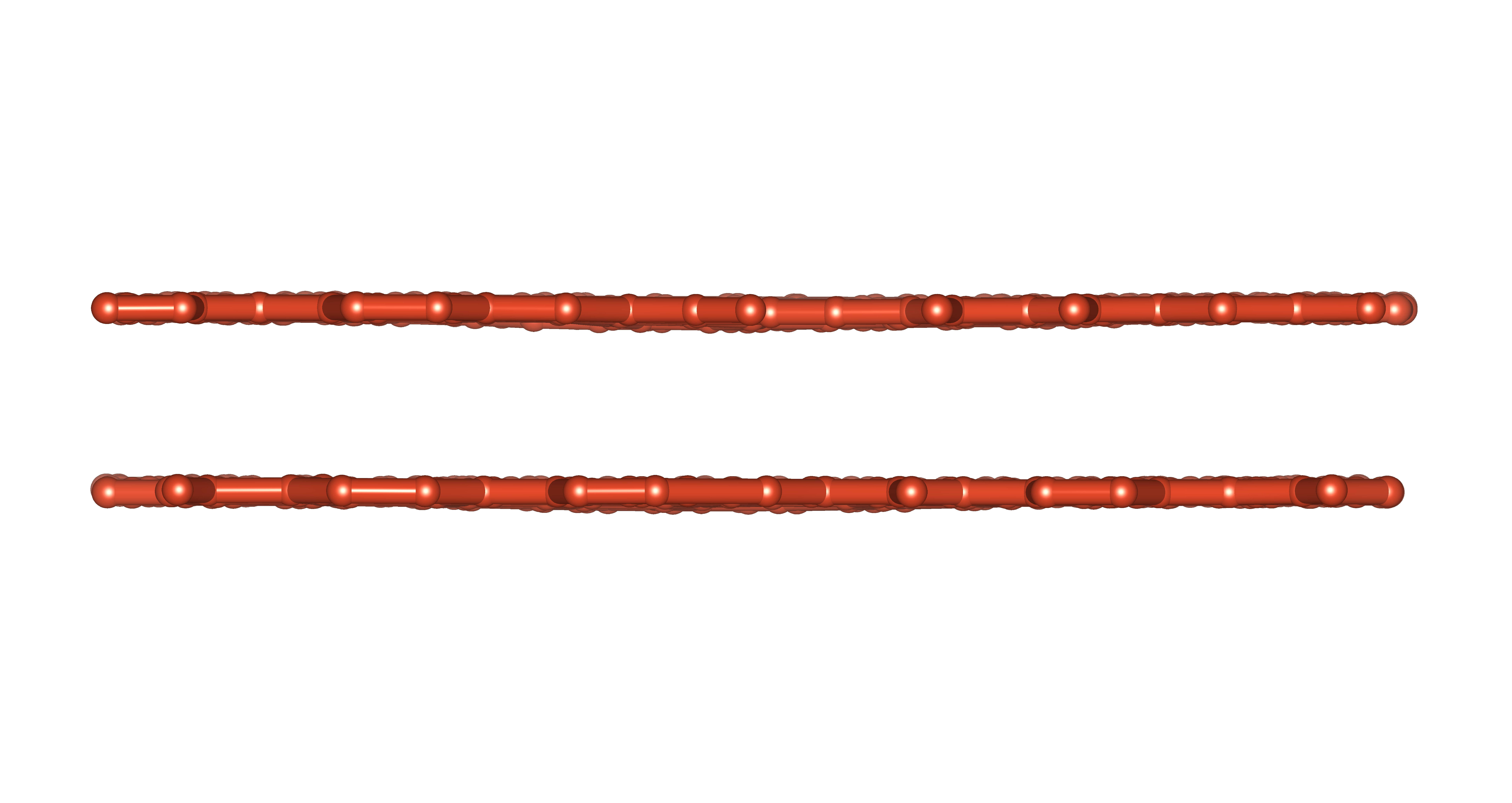} & & \\

%\hline
%\hline
%\hline
%Tri-layer & -6.01 & -91.86 & -46.88 \\
%\hline
\hhline{|~|-|-|-|-|-|}
%\cline{2-6}
& \multicolumn{2}{c|}{\multirow{3}{*}{Tri-layer}} & 20.04 & 459.31 & 234.38 \\
& \multicolumn{2}{c|}{} & \includegraphics[width=0.9in]{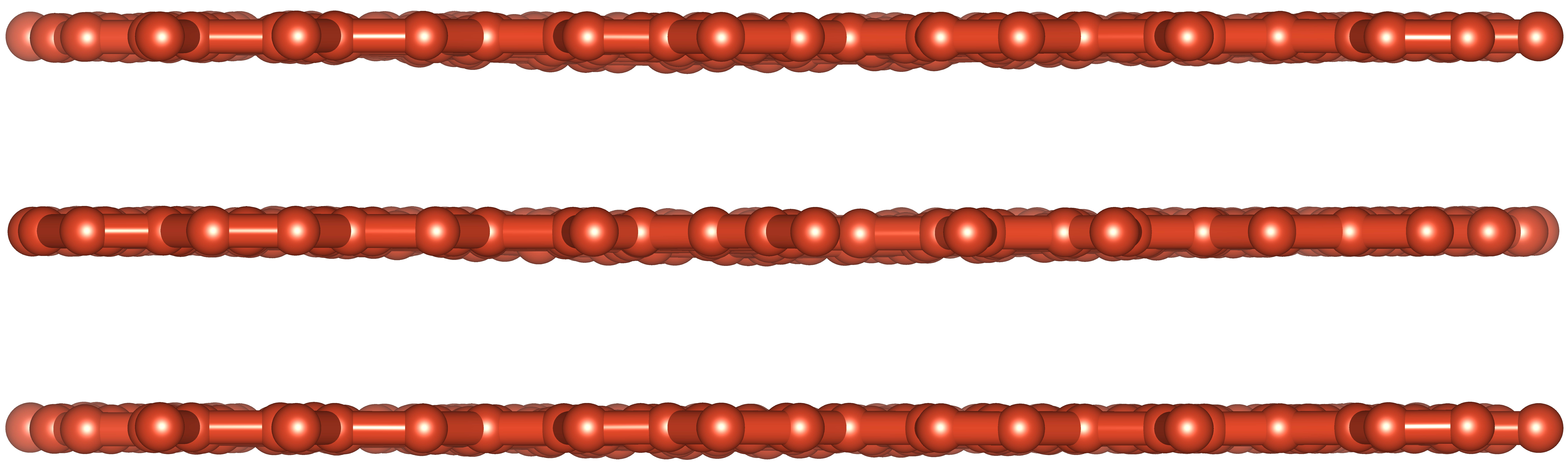}  &  \includegraphics[width=1.0in]{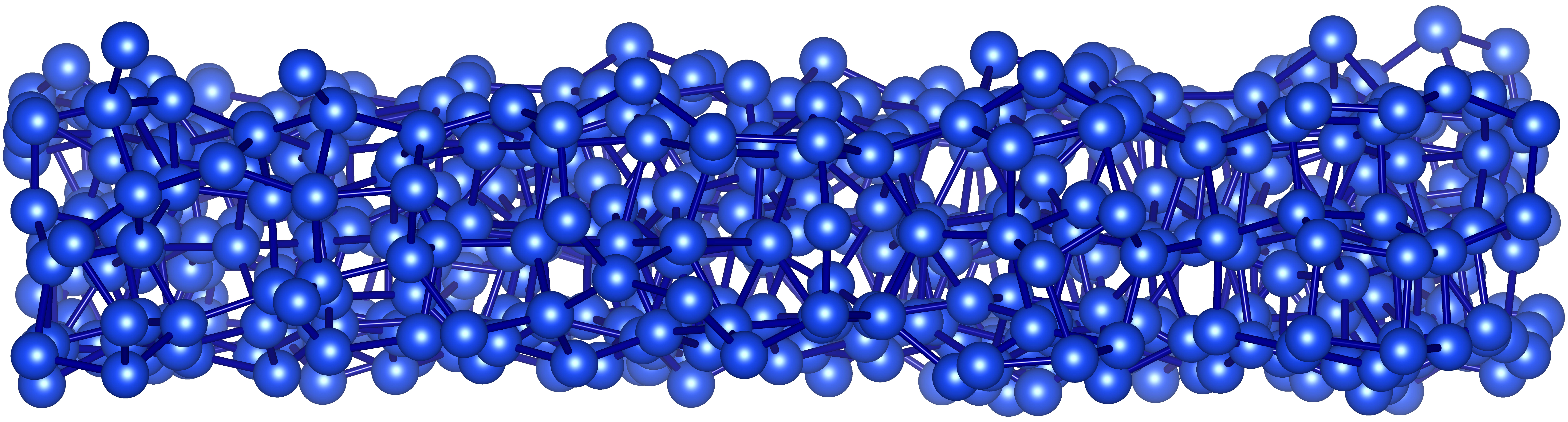} & \includegraphics[width=0.95in]{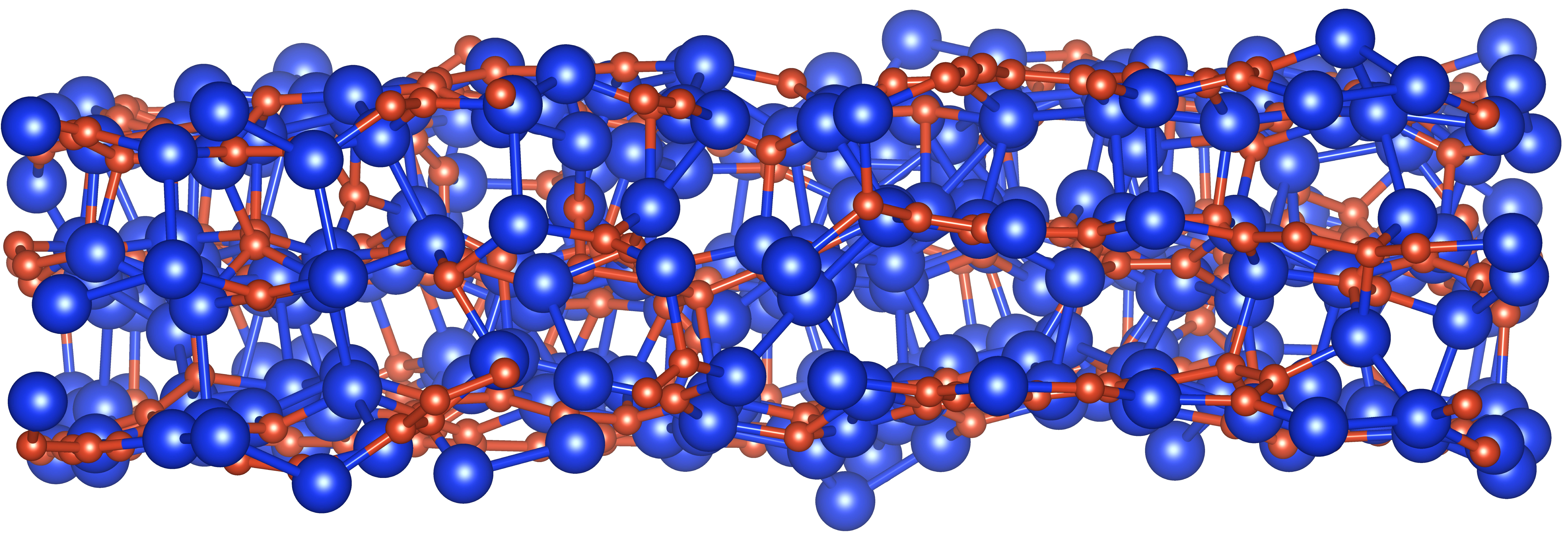} \\
%\hline
\hline

\end{tabular}
\caption{Calculated interlayer binding energies between A-Gra, A-Si and A-SiC layers in the multi layered systems. Double layer structures were classified in two groups; identical, referring to the case that the same layer being top of each other, and rotated, referring to the case the second layer being rotated by 90$^{\circ}$. The energetically more favourable structures' side view were appended below the calculated energies of that specific double layer systems. Tri-layer structures was modelled by keeping up and bottom layers same and only rotating the mid layers by 90 $^{\circ}$ cases.}
\label{table1}
\end{center}
\end{table*}

\subsubsection{Optical Properties}

\begin{figure*}
%\begin{center}
\centering
\includegraphics[width=\textwidth]{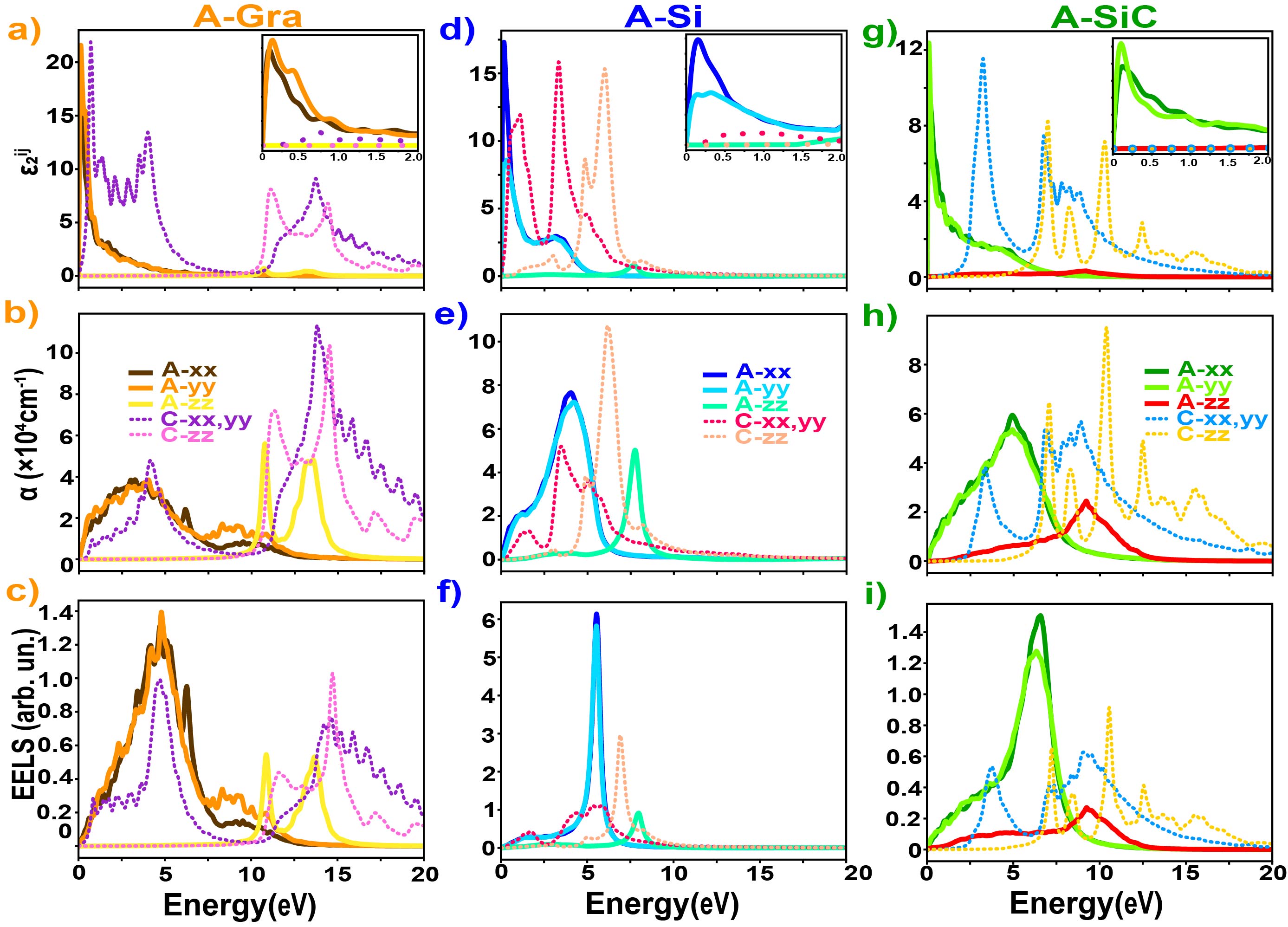}
\caption{Optical properties are represented for  A-Gra together with crystalline graphene; A-Si together with crystalline silicene; A-SiC together with crystalline silicon carbide. First row figures (\textbf{a}, \textbf{d}, \textbf{g}) show the calculated imaginary part of the dielectric functions (absorption spectrum with RPA and without LFE) for perpendicular (in plane: A-xx; A-yy; C-xx,yy) and parallel (out of plane: A-zz; C; ZZ) light polarizations  where the crystalline counterparts (C-xx,yy; C-zz) represented with the respectively  8; 6 and 6 times multiplied (only for dielectric function plot) for making a comparable visualization between amorphous and crystalline structures. Insets in the first row of figures show the spectrum form 0-2 eV without multiplication of crystalline systems. Second row figures (\textbf{b}, \textbf{e}, \textbf{h}) represent absorption coefficients. Third-row figures (\textbf{c}, \textbf{f}, \textbf{i}) represent the EELS spectrum of the systems.}
\label{fig4}
%\end{center}
\end{figure*}

Optical properties of $10\times10$ A-Gra, A-Si and A-SiC were studied and compared to their crystalline counterparts. We first focused on the imaginary part of dielectric functions, which is known as absorption spectrum Fig. \ref{fig4}(\textbf{a}, \textbf{d}, \textbf{g}), without LFE. The graphene absorption spectrum shows two main distinct peaks between 1-4 eV ($\pi \rightarrow \pi^{*}$ transition) and 10-14 eV ($\sigma \rightarrow \sigma^{*}$ transition) in perpendicular polarization. For the perpendicular polarization, the spectrum without LFE is almost similar to the one without LFE according to Ref. \onlinecite{Rubio}. The parallel polarization spectrum of graphene has two significant transitions at around 11 eV ($\pi \rightarrow \sigma^{*}$ transition) and 14 eV ($\sigma \rightarrow \pi^{*}$ transition). \cite{Rubio} It was shown that the LFE contribution has a significant effect on the spectrum, especially in the parallel polarization spectrum. As calculations including LFE for amorphous systems are numerically heavy, we present the comparison between amorphous and crystalline systems without considering LFE. Further, in cases of defected systems or in presence of adatoms, the spectrum shows a red-shift and a decrease in the peak intensities \cite{PentaGra-Optic,Gra-Gedop,Gra-SiC}. As in Fig. \ref{fig4} \textbf{a}, the peaks seen in the perpendicular polarization case for graphene are look-alike shifted and merged to a single large peak which is located at around 0.09 eV  along the x direction and 0.10 eV along the y direction, which are in IR range in the electromagnetic spectrum. However, graphene's peaks at 4 and 14  eV, exist in A-Gra's absorption spectrum at 5 eV (blue-shifted) and 9 eV (red-shifted), which are in the UV range in the electromagnetic spectrum. The static dielectric constant, which is $\epsilon_{1}(0)$, is found as in perpendicular polarization 40.41 along the x direction, 42.01 along the y direction, and  1.07 along parallel polarization. The static dielectric constant of crystalline graphene is found as 3.58 in the perpendicular polarization  and 1.21 in the parallel polarization. In-plane, the static dielectric constant of A-Gra found as 11.74 times larger while in-plane polarization was found as slightly lowered. Perpendicular polarization along the x and y direction spectrum could be accepted as close for the amorphous case, similar to the crystalline case. However, the absorption coefficient spectrum and EELS spectrum of both crystalline and amorphous graphene showed peaks at 4 and 14 eV. We found that the first plasmon peak of the amorphous structure occurs as a broader peak at the same frequency as the crystalline structure whereas the second plasmon peak of the amorphous structure is shifted as a broader peak to around 9 eV with lower intensity. The parallel polarization absorption spectrum of A-Gra shows the highest intensity at 13 eV as a merged broadened peak. The absorption coefficient and EELS spectrum showed the plasmon peaks of crystalline structure as red-shifted to 11 and 13 eV. Static refractive indexes of A-Gra are calculated as in perpendicular polarization 6.36 along the x direction and 6.49 along y direction, whereas 1.04 in parallel polarization. A-Gra's static refractive index was found 3.44 times larger than the crystalline graphene refractive index which was calculated as 1.89 in perpendicular polarization and 1.10 in parallel polarization.

Unlike the graphene case, A-Si absorption spectrum peaks were more visible, less merged, and more distinguishable. We found A-Si's absorption peaks as three peaks in perpendicular polarization, which are at respectively around 0.16 eV (IR); 2.41 eV (Visible-green); and 3.17 eV (Visible-violet), and as a single broad peak at around 7.67 eV (UV) in parallel polarization. In the crystalline silicene perpendicular polarization case, we found 3 peaks, at around 1.1 eV (IR); 3.3 eV (UV); and 4.93 eV (UV). While found 3 peaks in the parallel polarization case were at 3 eV (Visible-violet); 5 eV (UV); and 6 eV (UV). Whereas Mohan et al. found 3 peaks in perpendicular polarization and 3 peaks in parallel polarization at smaller energies using DFPT compare to our case. \cite{Siopt} Static dielectric constant of A-Si was calculated as in perpendicular polarization 33.56 along the x direction and 21.18 along the y direction, whereas 1.20 in parallel polarization direction, whereas crystalline silicene's static dielectric function in perpendicular and parallel polarization was found as respectively 3.88 and 1.57. A-Si static dielectric constant was found 8.65 times larger in perpendicular polarization, while 1.31 times larger in parallel polarization than crystalline silicene. A-Si's perpendicular polarization absorption peaks showed a red-shift compared to crystalline silicene, whereas parallel polarization components showed a single blue-shifted peak. Our calculated absorption peaks in perpendicular polarization for A-Si two merged peaks at 1.25 eV (IR) and 7.6 eV (UV), whereas crystalline showed the first peak at around 1.23 eV and the second peak appeared as a double peak at 3.49 and 5.1 eV. A-Si's parallel polarization absorption coefficient is located at 7.76 eV, while silicene's parallel polarization absorption coefficient peaks located at 6.17  eV (UV) with merged and relatively smaller peaks at 4.90 and 8.17 eV. With respect to the crystalline silicene's highest intensity parallel polarization peak, A-Si's showed a blue-shifted single peak similar to the absorption spectrum. In-plane plasmon peaks of A-Si were found located as a sharp peak at 5.54 eV in the UV range. Crystalline silicene's first peak appeared at 1.73 eV (near-IR) which has no such peak in A-Si, whereas the second peak appeared as a merged double peak but at around 5.3 eV. A-Si's plasmon peak has a sharp peak which is 6 times higher than crystalline silicene's second plasmon peak. In Parallel polarization, plasmon peaks were seen as sharp peaks in both cases. Where the A-Si's peak at 7.76 eV was seen as blue-shifted with smaller intensity compared to the crystalline's peak which is at around 6.9 eV. The static refractive index of A-Si is calculated as in perpendicular polarization along 5.79 along the x direction, 4.60 along the y direction, and 1.10 along the parallel polarization direction. Crystalline silicene's static refractive index in perpendicular and parallel polarization as respectively 1.97 and 1.25. A-Si's static refractive index was found 3 times larger in perpendicular polarization than crystalline, whereas in-plane polarization static refractive index was found slightly decreased relative to crystalline.
The a-SiC absorption spectrum was found as red-shifted relative to its crystalline counterpart. A-SiC absorption peaks were calculated at around 0.13 eV (IR), 3.54 eV (UV). While crystalline silicon carbide showed the first perpendicular polarization absorption peak at around 3.25 eV, the second peak was found as a large merged peak from 6.75 eV to 8.86 eV which has a good agreement with Shahrokhi's result\cite{Gra-SiC}. The perpendicular polarization of the absorption spectrum visibly shows a red-shift in A-SiC's spectrum relative to crystalline silicon carbide. Parallelly polarized absorption peaks of A-SiC located at 9.19 eV as a single and relatively narrower peak, whereas crystalline showed consecutive peaks  around a range between 7 to 17 eV. A-SiC's static dielectric constant is calculated as in perpendicular polarization 22.13 eV along the x direction, 24.68 eV along the y direction, and 1.27 eV along the parallel polarization direction. In crystalline silicon carbide, we found a static dielectric constant of 1.79 in perpendicular polarization and 1.26 in parallel polarization. A-SiC's static dielectric constant was found less than A-Si in perpendicular polarization, but 13.79 times larger than its crystalline counterpart. Absorption coefficients of A-SiC were found as a single broad merged peak where the highest intensity absorption occurred at 4.92 eV (UV), whereas crystalline absorption coefficient kept its first and second peaks, which were found in the absorption spectrum, as almost at the same energy range. Parallel polarization components of absorption coefficients showed the found peaks' maximums were kept at around the same frequencies in the UV range. A-SiC's parallel polarization component showed a broader peak than the absorption spectrum at 9.22 eV, however, different from the absorption spectrum the intensity decreases with respect to crystalline which was the same in both A-Gra and A-Si. The sharpest plasmon peak of A-SiC in-plane was found at 6.58 eV similar to A-Si's single sharp peak, whereas the range from 0 to 2 eV showed similar steeping behavior to A-Gra. Crystalline silicon carbide's two separate peaks from the absorption coefficient spectrum were  found as kept around the same energies in the plasmon spectrum. The parallel polarization plasmon peak of A-SiC was found at 9.45 eV, while crystalline silicon carbide was found as kept the consecutive transitions at around 6 eV to 17 eV. The static refractive index of A-SiC was found as in-plane 4.70 along the x direction, 4.97 along the y direction, and 1.13 along 1.13. Crystalline silicon carbide's static refractive index in perpendicular and parallel polarization was found as respectively 1.34 and 1.12. A-SiC's static refractive index in perpendicular polarization was found to decrease relative to A-Gra and A-Si, which is as expected relatively similar to the 50 \% Si-doped crystalline graphene system. The increasing static refractive index for A-Gra, A-Si, and A-SiC shows that these systems' absorption range and capacity are larger than crystalline counterparts and make them more advantageous for optoelectronic devices than crystalline alternatives. %Another found important finding about A-Gra, A-Si, and A-SiC's high static dielectric constant makes them a candidate for pyroelectric materials.????*

\subsection{Thermal Conductivity}\label{sec23}
We studied thermal conductivity with (50$\times$50) big cells. The previous studies show that thermal conductivity depends on amorphicity of the structure \cite{2DM-Agra,Nano-Agra}. Antidormi et al. \cite{2DM-Agra} showed that thermal conductivity decreases relative to how fast quenching is performed in the production step. The slower quench can induce an increase in the occurrence of the  probability of crystallization centers and local symmetry. The quicker production is done, the more glassy system production occurs. Therefore in agreement with Antidormi \cite{2DM-Agra,Nano-Agra}, the decreasing thermal conductivity with defect sites increasing in Pristine graphene, shows a decreasing thermal conductivity trend \cite{Nano-Agra}. In this work, we chose one fixed quenched rate and calculated the thermal conductivity of A-Gra; A-Si and A-SiC. We calculated thermal conductivity for A-Gra as $55.30 \pm 11.01$ W/Km; A-Si as $2.68 \pm 0.59$ W/Km; and A-SiC as $70.29 \pm 12.03$ W/Km. Compared to the previous studies, the found A-Gra thermal conductivity shows an agreement with the range in literature. It shows a significant difference with pristine graphene thermal conductivity which is in the interval 2600–3050 W/(Km) \cite{2DM-Agra}. A-Si thermal conductivity is interestingly showing a similar value to the calculated bulk amorphous Silicon's thermal conductivity \cite{Si2,Si3}. Zhou et al. also calculated amorphous Si-Nanowire as $2.4 \pm 0.25$ W/Km \cite{Si2}. This value compared to their calculated amorphous bulk Si is decreasing with dimension. Regarding the structural quality difference between their reference work and our study, our calculated value shows a good agreement with the literature. Compared to our studied A-Gra and A-Si, we calculated A-SiC's thermal conductivity higher than A-Gra and A-Si. Regarding A-Gra's and A-SiC's calculated statistical error range, this result can be accepted. 
We calculated A-SiC's thermal conductivity as $ 70.28 \pm 12.03 $ W/Km. We can see that A-SiC's thermal conductivity dominantly originated from its C atoms more than silicon atoms. In accordance with the selected fast-quenching parameters, the produced A-Gra shows a higher thermal conductivity compared to the produced A-Si by using the defined parameters. As a result, the structural quality can be a source of difference between the calculated relatively  higher thermal conductivity than the existing literature values. Our work debuts that the carbon webbing in A-SiC, which is in 2D, plays a decisive role in its thermal conductivity.

\subsection{Vibrational Properties}\label{sec24}
\begin{figure*}
\centering
\includegraphics[width=1.0\textwidth]{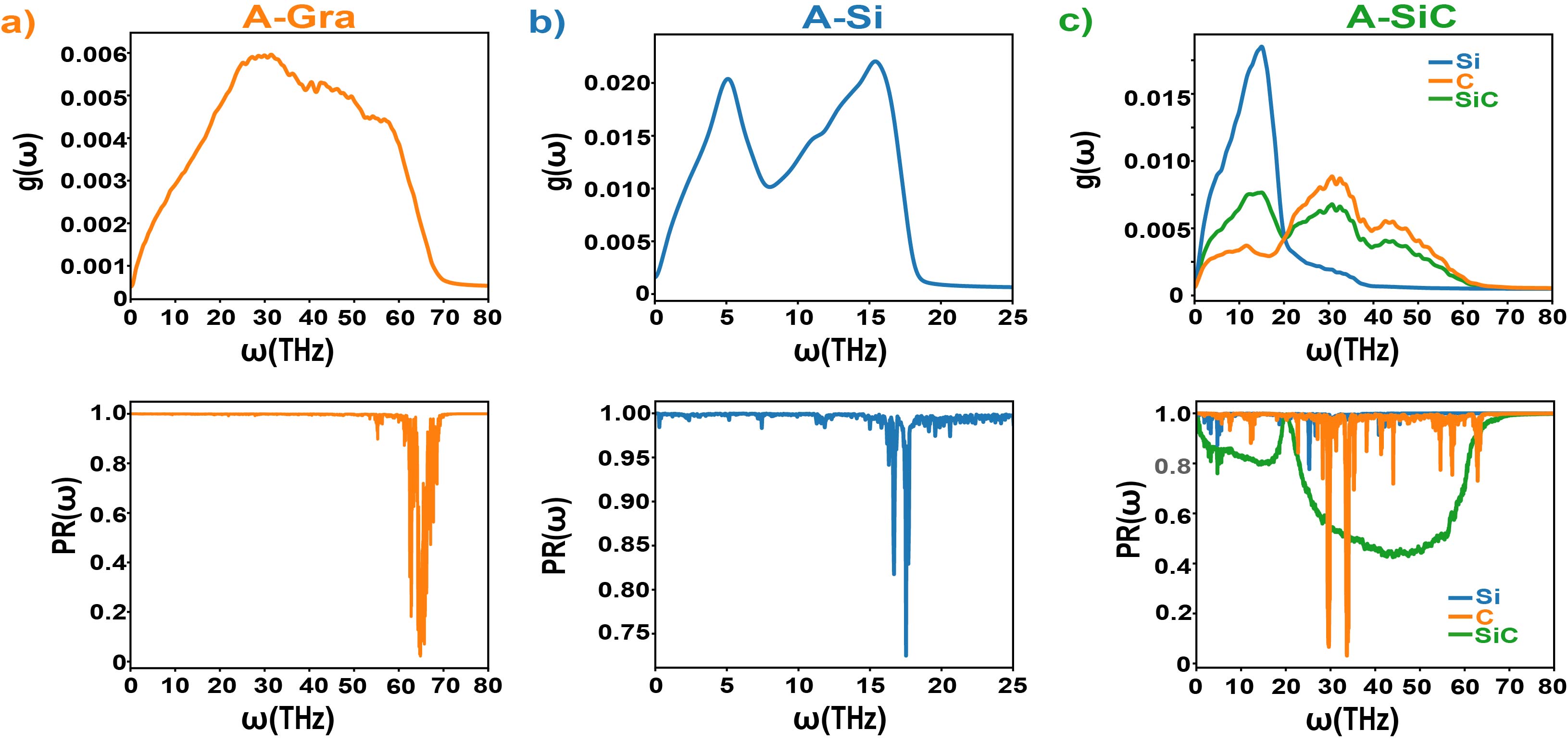}
\caption{VDOS (on the top row) and PR (on the bottom row) plots of the systems' is represented respectively a) A-Gra; b) A-Si and c) A-SiC(together with total and atom projected vdos and PR).}
\label{fig5}
\end{figure*}
We made a further investigation of calculated low thermal conductivity of $50\times50$ A-Gra, A-Si and A-SiC by calculating the vibrational density of states (VDOS) and mode participation ratios (PR). As a known fact that amorphous structures' VDOS show broader merged peaks than their crystalline counterparts. Amorphous structures' vibrational properties show similar trends to 3D-bulk cases with decreasing dimension \cite{Giri-1,2DM-2El,Si-SiC,Si2,Si3,SiC-thermal}, which we can see in our work Fig. \ref{fig5}. The density can change these similarities by mode intensity increase/decrease and mode frequency shifts relative to the new webbing and hybridization of the constituting atoms\cite{Giri-1}.  Our calculated VDOS for A-Gra show similar broadening relative to crystalline counterparts and to the literature data in different densities; structural quality  \cite{2DM-Agra,Nano-Agra,2DM-2El,Carbonmech-Agra}. Though we cannot quantify the extendon (propagons, diffusion) and locon modes, because Green-Kubo method gives equipartition to all modes, extendon modes are the dominant heat carriers. A-Gra has high-intensity low frequency extendon modes compared to pristine graphene. The highest intensity having extendon mode is between $ \sim 25 - 35 $ THz. The locon modes start at around $ \sim 60$ THz. Most of the literature data show the activation of the locon mode at around $ \sim 50 $ THz. This is actually relative to the system's amorphousness and density. Denser and highly amorphous systems can show a redshift in locon modes. Our system has high intensity of extendons below $ \sim 50 $ THz, which can be explained by the expectation of increasing \textit{sp$^3$} hybridization states or the structural high density. Here the studied cases are neither at low nor at high density. The structurally denser systems are expected to show higher \textit{sp$^3$} hybridization states similar to diamond, so the phonon-like heat carriers in the glassy systems, which are known as propagon modes increase and this increase gives rise in the system's thermal conductivity. We claim that structural quality depending on cooling rate can increase the \textit{sp$^3$} hybridization states. And therefore our A-Gra system shows $\sim 70 $W/Km thermal conductivity and this becomes visible in VDOS and PR as an increase in extendon modes. Without applying an Allen-Fellman modal analysis, it is hard to discriminate between the propagon, diffusion and locon modes' transition frequencies. Though A-Gra's calculated PR shows phonon-like contribution ($ PR = 1 $) until locon modes' sharp decrease ($ PR = 0.2 $), we know this is coming from EMD-based calculation, where all the modes' contribution is included. Moreover, the  regime especially in the range 10 THz-50THz is dominantly diffusions' regime and this can be interpreted as that the diffusions are the important heat carriers in the system in addition to the propagons. However, extendons' increase is enough to explain A-Gra thermal conductivity. Giri et al \cite{Giri-1}'s made VDOS analysis of bulk amorphous carbon with the density of 2.1 gcm$^-3$, which is close to our density, and their mode frequency regime looks similar to ours. However, the structural quality depends on the chosen parameters and in connection to this  the observed carbon webbing difference and dimensionality decrease is responsible for the difference in thermal conductivity by showing similar VDOS. This is similar to our A-Si case. A-Si VDOS is comparable to the literature VDOS of bulk amorphous silicon in similar density. A Denser system can decrease the first peak intensity, and increase the second peak intensity, and this can cause a blue-shift in the tail that is being seen as locon frequencies. However PR is different from the literature, our system is not showing a locon peak as defined in the literature for bulk cases. We assume this can be sourced from the dimension decrease. This is an interesting finding that the system shows low thermal conductivity but lacks of locon modes, which agrees with the reference \cite{Si3} where they found amorphous silicene's dominant heat carriers as propagons. Though our interpretation of diffusion modes increases with decreasing dimension, a Green-Kubo Modal Analysis can be applied as further study to quantitatively report the increase.
A-SiC's calculated VDOS carries the traces of both A-Gra and A-Si while the bonding nature and density for C and Si atoms are different from A-Gra and A-Si. The difference is visible as sharp red-shifts in locon modes ($\sim 30$ THz) of C atoms in the partial VDOS Fig. \ref{fig5}. A similar red-Shift is also visible in Si atoms diffusion modes(beyond $\sim 20$ THz). The overall VDOS shows a reasonable agreement with Li et al.'s amorphous bulk SiC VDOS plot with a similar density. Through the plotted atom projected VDOS', Si atoms contribute to the overall A-SiC VDOS plot as unified and red-shifted extendon modes, while C atoms' contributions can be seen as increasing second peak intensity, however, red-shift on the frequency ranges of  the modes of both Si and C atoms' similarly exist on  the plots. C atoms' diffusion modes visibly increased to a wide range $ \sim 63 $ THz in the A-SiC webbing nature where the locon modes' frequency range narrows and shows forward shift   $\sim 30$ THz which is visible in the PR plot. However, C atoms' are showing locon modes on projected VDOS, overall A-SiC is not showing localized modes whereas it is showing a wider range for diffusion modes. 
We can conclude the section by emphasizing that at the calculated densities of the structures, the decreasing dimension decreases localized modes. The main heat carriers in the systems are  diffusions. This can explain the reason of A-Gra and A-SiC thermal conductivities' which are higher relative to A-Si.

\section{Conclusion}
In conclusion, we have made a detailed study of 2D A-Gra, A-Si, and A-SiC by atomistic simulations. Our simulated amorphous structures agree reasonably well with the literature, wherever available. A-Gra is found to be planar at 0 K while it gets wrinkled at 300 K. In this study, we made the A-SiC analysis for the first time. All three structures show unique bond lengths, angles and ring sizes unique to their constituent atoms, amorphicity, and density. 
We have also shown all the studied structures' electronic properties by using first principles. All three of the studied structures are metallic and show different charge distributions on local rings. This can be useful for future functional device designing where local functionalization is required in the system. In A-SiC, the charge density distribution around C and Si atoms are somewhat similar to their constituent's density but also carries some special features due to their coexistence.
We found A-Gra, A-Si and A-SiC's optical properties comparable to the crystalline counterparts. Except for A-Si's parallelly polarized components, all calculated spectra for all three structures showed red-shift compared to their crystalline counterparts. All three structures were found as absorbing from IR to UV range and plasmon frequencies in the UV range. 
We also showed that 2D A-Gra structures can create vdW solids by keeping their electronic properties. While A-Si prefers to create covalent bonding between the layers. One interesting finding is seen in A-SiC layered structure as it partially favors creating bonds between Si atoms in contrast to not bonding C atoms.   

Furthermore, we calculated thermal conductivity by using the Green-Kubo method. The calculated value for A-Gra as 55.30 W/K.m is in agreement with some articles in the literature. However, the value of thermal conductivity depends critically on the density and parameters used to generate the amorphous structures. In contrast to A-gra, A-Si shows a low thermal conductivity, 2.68 W/K.m, as found in bulk amorphous systems and 1D amorphous nanowires. Finally, A-SiC's thermal conductivity is found to be 70.29 W/K.m, which has the highest value among the three systems studied.

From the analysis of VDOS and PR, we find that the extendons are the primary carriers of thermal transport in all the three systems. However, we have also observed distinct localized modes in A-Gra in the frequency range of 60-70 THz. The red-shifts in the mode frequency range relative to literature values have been discussed in detail which depends on the structures' density and the parameters used in the production step. One important finding is that A-SiC's carbon atoms show localized modes in the atom projected DOS and PR while this is not seen on the overall A-SiC. The dominance of extendons is thought to be the explanation of A-Gra and similarly A-SiC's relatively high thermal conductivity. We think the decreasing dimension has a visible increase in  heat carrier diffusions in contrast to the localized modes. 2D amorphous structures  can be seen as an alternative to future electronic device design materials With these specific and tuneable electronic and thermal properties. 

\section{Acknowledgements}
B.S. acknowledges financial support from Swedish Research Council (grant no. 2022-04309). The computations were enabled in project SNIC 2022/3-30 by resources provided by the Swedish National Infrastructure for Computing (SNIC) at NSC, PDC, and HPC2N partially funded by the Swedish Research Council (Grant No. 2018-05973). B.S. also acknowledges allocation of supercomputing hours by PRACE DECI-17 project 'Q2Dtopomat' in Eagle supercomputer in Poland and EuroHPC resources in Karolina supercomputer in the Czech Republic.

%%%%%%%\bibliographystyle{plain}        % Include this if you use bibtex 
%\addbibresource{sample.bib}
%\bibliography{sample} 
%\bibliographystyle{unsrt} 
\bibliographystyle{apsrev4-2}
\bibliography{autosam}

\providecommand{\noopsort}[1]{}\providecommand{\singleletter}[1]{#1}%
\begin{thebibliography}{58}%
\makeatletter
\providecommand \@ifxundefined [1]{%
 \@ifx{#1\undefined}
}%
\providecommand \@ifnum [1]{%
 \ifnum #1\expandafter \@firstoftwo
 \else \expandafter \@secondoftwo
 \fi
}%
\providecommand \@ifx [1]{%
 \ifx #1\expandafter \@firstoftwo
 \else \expandafter \@secondoftwo
 \fi
}%
\providecommand \natexlab [1]{#1}%
\providecommand \enquote  [1]{``#1''}%
\providecommand \bibnamefont  [1]{#1}%
\providecommand \bibfnamefont [1]{#1}%
\providecommand \citenamefont [1]{#1}%
\providecommand \href@noop [0]{\@secondoftwo}%
\providecommand \href [0]{\begingroup \@sanitize@url \@href}%
\providecommand \@href[1]{\@@startlink{#1}\@@href}%
\providecommand \@@href[1]{\endgroup#1\@@endlink}%
\providecommand \@sanitize@url [0]{\catcode `\\12\catcode `\$12\catcode
  `\&12\catcode `\#12\catcode `\^12\catcode `\_12\catcode `\%12\relax}%
\providecommand \@@startlink[1]{}%
\providecommand \@@endlink[0]{}%
\providecommand \url  [0]{\begingroup\@sanitize@url \@url }%
\providecommand \@url [1]{\endgroup\@href {#1}{\urlprefix }}%
\providecommand \urlprefix  [0]{URL }%
\providecommand \Eprint [0]{\href }%
\providecommand \doibase [0]{https://doi.org/}%
\providecommand \selectlanguage [0]{\@gobble}%
\providecommand \bibinfo  [0]{\@secondoftwo}%
\providecommand \bibfield  [0]{\@secondoftwo}%
\providecommand \translation [1]{[#1]}%
\providecommand \BibitemOpen [0]{}%
\providecommand \bibitemStop [0]{}%
\providecommand \bibitemNoStop [0]{.\EOS\space}%
\providecommand \EOS [0]{\spacefactor3000\relax}%
\providecommand \BibitemShut  [1]{\csname bibitem#1\endcsname}%
\let\auto@bib@innerbib\@empty
%</preamble>
\bibitem [{\citenamefont {Street}(2000)}]{Am-general}%
  \BibitemOpen
  \bibfield  {author} {\bibinfo {author} {\bibfnamefont {R.~A.}\ \bibnamefont
  {Street}},\ }\href@noop {} {\bibfield  {journal} {\bibinfo  {journal}
  {Springer}\ } (\bibinfo {year} {2000})}\BibitemShut {NoStop}%
\bibitem [{\citenamefont {Joo}\ \emph {et~al.}(2017)\citenamefont {Joo},
  \citenamefont {Lee}, \citenamefont {Jang}, \citenamefont {Kang},
  \citenamefont {Kwon}, \citenamefont {Chung}, \citenamefont {Lee},
  \citenamefont {Kim}, \citenamefont {Kim}, \citenamefont {Yang}, \citenamefont
  {Kim}, \citenamefont {Choi}, \citenamefont {Whang},\ and\ \citenamefont
  {Hwang}}]{Zachariasen}%
  \BibitemOpen
  \bibfield  {author} {\bibinfo {author} {\bibfnamefont {W.-J.}\ \bibnamefont
  {Joo}}, \bibinfo {author} {\bibfnamefont {J.-H.}\ \bibnamefont {Lee}},
  \bibinfo {author} {\bibfnamefont {Y.}~\bibnamefont {Jang}}, \bibinfo {author}
  {\bibfnamefont {S.-G.}\ \bibnamefont {Kang}}, \bibinfo {author}
  {\bibfnamefont {Y.-N.}\ \bibnamefont {Kwon}}, \bibinfo {author}
  {\bibfnamefont {J.}~\bibnamefont {Chung}}, \bibinfo {author} {\bibfnamefont
  {S.}~\bibnamefont {Lee}}, \bibinfo {author} {\bibfnamefont {C.}~\bibnamefont
  {Kim}}, \bibinfo {author} {\bibfnamefont {T.-H.}\ \bibnamefont {Kim}},
  \bibinfo {author} {\bibfnamefont {C.-W.}\ \bibnamefont {Yang}}, \bibinfo
  {author} {\bibfnamefont {U.~J.}\ \bibnamefont {Kim}}, \bibinfo {author}
  {\bibfnamefont {B.~L.}\ \bibnamefont {Choi}}, \bibinfo {author}
  {\bibfnamefont {D.}~\bibnamefont {Whang}},\ and\ \bibinfo {author}
  {\bibfnamefont {S.-W.}\ \bibnamefont {Hwang}},\ }\href
  {https://doi.org/10.1126/sciadv.1601821} {\bibfield  {journal} {\bibinfo
  {journal} {Sci. Adv.}\ }\textbf {\bibinfo {volume} {3}},\ \bibinfo {pages}
  {1601821} (\bibinfo {year} {2017})}\BibitemShut {NoStop}%
\bibitem [{\citenamefont {Zhou}\ \emph {et~al.}(2020)\citenamefont {Zhou},
  \citenamefont {Cheng}, \citenamefont {Chen}, \citenamefont {Xie},
  \citenamefont {Wang},\ and\ \citenamefont {Zhang}}]{all}%
  \BibitemOpen
  \bibfield  {author} {\bibinfo {author} {\bibfnamefont {W.}~\bibnamefont
  {Zhou}}, \bibinfo {author} {\bibfnamefont {Y.}~\bibnamefont {Cheng}},
  \bibinfo {author} {\bibfnamefont {K.}~\bibnamefont {Chen}}, \bibinfo {author}
  {\bibfnamefont {G.}~\bibnamefont {Xie}}, \bibinfo {author} {\bibfnamefont
  {T.}~\bibnamefont {Wang}},\ and\ \bibinfo {author} {\bibfnamefont
  {G.}~\bibnamefont {Zhang}},\ }\href
  {https://doi.org/https://doi.org/10.1002/adfm.201903829} {\bibfield
  {journal} {\bibinfo  {journal} {Adv. Func. Mater.}\ }\textbf {\bibinfo
  {volume} {30}},\ \bibinfo {pages} {1903829} (\bibinfo {year}
  {2020})}\BibitemShut {NoStop}%
\bibitem [{\citenamefont {Li}\ \emph {et~al.}(2022)\citenamefont {Li},
  \citenamefont {Zheng}, \citenamefont {Cao}, \citenamefont {Zhang},\ and\
  \citenamefont {Xia}}]{2DA-Ex-GraTrend}%
  \BibitemOpen
  \bibfield  {author} {\bibinfo {author} {\bibfnamefont {C.}~\bibnamefont
  {Li}}, \bibinfo {author} {\bibfnamefont {C.}~\bibnamefont {Zheng}}, \bibinfo
  {author} {\bibfnamefont {F.}~\bibnamefont {Cao}}, \bibinfo {author}
  {\bibfnamefont {Y.}~\bibnamefont {Zhang}},\ and\ \bibinfo {author}
  {\bibfnamefont {X.}~\bibnamefont {Xia}},\ }\href
  {https://doi.org/10.1007/s11664-022-09687-4} {\bibfield  {journal} {\bibinfo
  {journal} {J. Electron. Mater.}\ }\textbf {\bibinfo {volume} {51}},\ \bibinfo
  {pages} {4107–4114} (\bibinfo {year} {2022})}\BibitemShut {NoStop}%
\bibitem [{\citenamefont {Butler}\ \emph {et~al.}(2013)\citenamefont {Butler},
  \citenamefont {Hollen}, \citenamefont {Cao}, \citenamefont {Cui},
  \citenamefont {Gupta}, \citenamefont {Gutiérrez}, \citenamefont {Heinz},
  \citenamefont {Hong}, \citenamefont {Huang}, \citenamefont {Ismach},
  \citenamefont {Johnston-Halperin}, \citenamefont {Kuno}, \citenamefont
  {Plashnitsa}, \citenamefont {Robinson}, \citenamefont {Ruoff}, \citenamefont
  {Salahuddin}, \citenamefont {Shan}, \citenamefont {Shi}, \citenamefont
  {Spencer}, \citenamefont {Terrones}, \citenamefont {Windl},\ and\
  \citenamefont {Goldberger}}]{2DCrystaloverview}%
  \BibitemOpen
  \bibfield  {author} {\bibinfo {author} {\bibfnamefont {S.~Z.}\ \bibnamefont
  {Butler}}, \bibinfo {author} {\bibfnamefont {S.~M.}\ \bibnamefont {Hollen}},
  \bibinfo {author} {\bibfnamefont {L.}~\bibnamefont {Cao}}, \bibinfo {author}
  {\bibfnamefont {Y.}~\bibnamefont {Cui}}, \bibinfo {author} {\bibfnamefont
  {J.~A.}\ \bibnamefont {Gupta}}, \bibinfo {author} {\bibfnamefont {H.~M.}\
  \bibnamefont {Gutiérrez}}, \bibinfo {author} {\bibfnamefont {T.~F.}\
  \bibnamefont {Heinz}}, \bibinfo {author} {\bibfnamefont {S.~S.}\ \bibnamefont
  {Hong}}, \bibinfo {author} {\bibfnamefont {J.}~\bibnamefont {Huang}},
  \bibinfo {author} {\bibfnamefont {A.~F.}\ \bibnamefont {Ismach}}, \bibinfo
  {author} {\bibfnamefont {E.}~\bibnamefont {Johnston-Halperin}}, \bibinfo
  {author} {\bibfnamefont {M.}~\bibnamefont {Kuno}}, \bibinfo {author}
  {\bibfnamefont {V.~V.}\ \bibnamefont {Plashnitsa}}, \bibinfo {author}
  {\bibfnamefont {R.~D.}\ \bibnamefont {Robinson}}, \bibinfo {author}
  {\bibfnamefont {R.~S.}\ \bibnamefont {Ruoff}}, \bibinfo {author}
  {\bibfnamefont {S.}~\bibnamefont {Salahuddin}}, \bibinfo {author}
  {\bibfnamefont {J.}~\bibnamefont {Shan}}, \bibinfo {author} {\bibfnamefont
  {L.}~\bibnamefont {Shi}}, \bibinfo {author} {\bibfnamefont {M.~G.}\
  \bibnamefont {Spencer}}, \bibinfo {author} {\bibfnamefont {M.}~\bibnamefont
  {Terrones}}, \bibinfo {author} {\bibfnamefont {W.}~\bibnamefont {Windl}},\
  and\ \bibinfo {author} {\bibfnamefont {J.~E.}\ \bibnamefont {Goldberger}},\
  }\href {https://doi.org/10.1021/nn400280c} {\bibfield  {journal} {\bibinfo
  {journal} {ACS Nano}\ }\textbf {\bibinfo {volume} {7}},\ \bibinfo {pages}
  {2898–2926} (\bibinfo {year} {2013})}\BibitemShut {NoStop}%
\bibitem [{\citenamefont {Zhang}\ \emph {et~al.}(2018)\citenamefont {Zhang},
  \citenamefont {Chhowalla},\ and\ \citenamefont {Liu}}]{2DCrystaloverview2}%
  \BibitemOpen
  \bibfield  {author} {\bibinfo {author} {\bibfnamefont {H.}~\bibnamefont
  {Zhang}}, \bibinfo {author} {\bibfnamefont {M.}~\bibnamefont {Chhowalla}},\
  and\ \bibinfo {author} {\bibfnamefont {Z.}~\bibnamefont {Liu}},\ }\href
  {https://doi.org/10.1039/C8CS90048E} {\bibfield  {journal} {\bibinfo
  {journal} {Chem. Soc. Rev.}\ }\textbf {\bibinfo {volume} {47}},\ \bibinfo
  {pages} {3015–3017} (\bibinfo {year} {2018})}\BibitemShut {NoStop}%
\bibitem [{\citenamefont {Novoselov}\ \emph {et~al.}(2016)\citenamefont
  {Novoselov}, \citenamefont {Mishcenko}, \citenamefont {Carvalho},\ and\
  \citenamefont {Castro~Neto}}]{2DvdW}%
  \BibitemOpen
  \bibfield  {author} {\bibinfo {author} {\bibfnamefont {K.~S.}\ \bibnamefont
  {Novoselov}}, \bibinfo {author} {\bibfnamefont {A.}~\bibnamefont
  {Mishcenko}}, \bibinfo {author} {\bibfnamefont {A.}~\bibnamefont
  {Carvalho}},\ and\ \bibinfo {author} {\bibfnamefont {A.~H.}\ \bibnamefont
  {Castro~Neto}},\ }\href {https://doi.org/10.1126/science.aac9439} {\bibfield
  {journal} {\bibinfo  {journal} {Science}\ }\textbf {\bibinfo {volume}
  {353}},\ \bibinfo {pages} {461} (\bibinfo {year} {2016})}\BibitemShut
  {NoStop}%
\bibitem [{\citenamefont {Duong}\ \emph {et~al.}(2017)\citenamefont {Duong},
  \citenamefont {Yun},\ and\ \citenamefont {Lee}}]{2DvdW1}%
  \BibitemOpen
  \bibfield  {author} {\bibinfo {author} {\bibfnamefont {D.~L.}\ \bibnamefont
  {Duong}}, \bibinfo {author} {\bibfnamefont {S.~J.}\ \bibnamefont {Yun}},\
  and\ \bibinfo {author} {\bibfnamefont {Y.~H.}\ \bibnamefont {Lee}},\ }\href
  {https://doi.org/10.1021/acsnano.7b07436} {\bibfield  {journal} {\bibinfo
  {journal} {ACS}\ }\textbf {\bibinfo {volume} {11}},\ \bibinfo {pages} {11803}
  (\bibinfo {year} {2017})}\BibitemShut {NoStop}%
\bibitem [{\citenamefont {Eder}\ \emph {et~al.}(2014)\citenamefont {Eder},
  \citenamefont {Kotakoski}, \citenamefont {Kaiser},\ and\ \citenamefont
  {Meyer}}]{Zachariasen-2}%
  \BibitemOpen
  \bibfield  {author} {\bibinfo {author} {\bibfnamefont {F.~R.}\ \bibnamefont
  {Eder}}, \bibinfo {author} {\bibfnamefont {J.}~\bibnamefont {Kotakoski}},
  \bibinfo {author} {\bibfnamefont {U.}~\bibnamefont {Kaiser}},\ and\ \bibinfo
  {author} {\bibfnamefont {J.~C.}\ \bibnamefont {Meyer}},\ }\href
  {https://doi.org/10.1038/srep04060} {\bibfield  {journal} {\bibinfo
  {journal} {Sci. Rep.}\ }\textbf {\bibinfo {volume} {4}},\ \bibinfo {pages}
  {4060} (\bibinfo {year} {2014})}\BibitemShut {NoStop}%
\bibitem [{\citenamefont {Zhao}\ \emph {et~al.}(2019)\citenamefont {Zhao},
  \citenamefont {Chen}, \citenamefont {Wang}, \citenamefont {Qui},\ and\
  \citenamefont {Guo}}]{2DA-Exp-IOP}%
  \BibitemOpen
  \bibfield  {author} {\bibinfo {author} {\bibfnamefont {H.}~\bibnamefont
  {Zhao}}, \bibinfo {author} {\bibfnamefont {X.}~\bibnamefont {Chen}}, \bibinfo
  {author} {\bibfnamefont {G.}~\bibnamefont {Wang}}, \bibinfo {author}
  {\bibfnamefont {Y.}~\bibnamefont {Qui}},\ and\ \bibinfo {author}
  {\bibfnamefont {L.}~\bibnamefont {Guo}},\ }\href
  {https://doi.org/10.1088/2053-1583/ab1169} {\bibfield  {journal} {\bibinfo
  {journal} {2DMater.}\ }\textbf {\bibinfo {volume} {6}},\ \bibinfo {pages}
  {032002} (\bibinfo {year} {2019})}\BibitemShut {NoStop}%
\bibitem [{\citenamefont {Yang}\ \emph {et~al.}(2020)\citenamefont {Yang},
  \citenamefont {Hao},\ and\ \citenamefont {Lau}}]{2DA-Ex-JAP}%
  \BibitemOpen
  \bibfield  {author} {\bibinfo {author} {\bibfnamefont {Z.}~\bibnamefont
  {Yang}}, \bibinfo {author} {\bibfnamefont {J.}~\bibnamefont {Hao}},\ and\
  \bibinfo {author} {\bibfnamefont {S.~P.}\ \bibnamefont {Lau}},\ }\href
  {https://doi.org/10.1063/1.5144626} {\bibfield  {journal} {\bibinfo
  {journal} {J. Appl. Phys.}\ }\textbf {\bibinfo {volume} {127}},\ \bibinfo
  {pages} {220901} (\bibinfo {year} {2020})}\BibitemShut {NoStop}%
\bibitem [{\citenamefont {He}\ \emph {et~al.}(2022)\citenamefont {He},
  \citenamefont {Liu}, \citenamefont {Zhu},\ and\ \citenamefont
  {et~al.}}]{2DA-Ex-Am-Chalcogen}%
  \BibitemOpen
  \bibfield  {author} {\bibinfo {author} {\bibfnamefont {Y.}~\bibnamefont
  {He}}, \bibinfo {author} {\bibfnamefont {L.}~\bibnamefont {Liu}}, \bibinfo
  {author} {\bibfnamefont {C.}~\bibnamefont {Zhu}},\ and\ \bibinfo {author}
  {\bibnamefont {et~al.}},\ }\href {https://doi.org/10.1038/s41929-022-00753-y}
  {\bibfield  {journal} {\bibinfo  {journal} {Nat. Catal.}\ }\textbf {\bibinfo
  {volume} {5}},\ \bibinfo {pages} {212} (\bibinfo {year} {2022})}\BibitemShut
  {NoStop}%
\bibitem [{\citenamefont {Pan}\ \emph {et~al.}(2014)\citenamefont {Pan},
  \citenamefont {Hinks}, \citenamefont {Ramasse}, \citenamefont {Greaves},
  \citenamefont {Bangert}, \citenamefont {Donnely},\ and\ \citenamefont
  {Haigh}}]{2DA-Ex-SciRep}%
  \BibitemOpen
  \bibfield  {author} {\bibinfo {author} {\bibfnamefont {C.~T.}\ \bibnamefont
  {Pan}}, \bibinfo {author} {\bibfnamefont {J.~A.}\ \bibnamefont {Hinks}},
  \bibinfo {author} {\bibfnamefont {Q.~M.}\ \bibnamefont {Ramasse}}, \bibinfo
  {author} {\bibfnamefont {G.}~\bibnamefont {Greaves}}, \bibinfo {author}
  {\bibfnamefont {U.}~\bibnamefont {Bangert}}, \bibinfo {author} {\bibfnamefont
  {S.~E.}\ \bibnamefont {Donnely}},\ and\ \bibinfo {author} {\bibfnamefont
  {S.~J.}\ \bibnamefont {Haigh}},\ }\href {https://doi.org/10.1038/srep06334}
  {\bibfield  {journal} {\bibinfo  {journal} {Sci. Rep.}\ }\textbf {\bibinfo
  {volume} {4}},\ \bibinfo {pages} {6334} (\bibinfo {year} {2014})}\BibitemShut
  {NoStop}%
\bibitem [{\citenamefont {Toh}\ \emph {et~al.}(2020)\citenamefont {Toh},
  \citenamefont {Zhang}, \citenamefont {Lin}, \citenamefont {Mayorov},
  \citenamefont {Wang}, \citenamefont {Orofeo}, \citenamefont {Ferry},
  \citenamefont {Andersen}, \citenamefont {Kakenov}, \citenamefont {Guo},
  \citenamefont {Abidi}, \citenamefont {Sims}, \citenamefont {Suenaga},
  \citenamefont {Pantelides},\ and\ \citenamefont {Özyilmaz
  B.}}]{AGra-Nature}%
  \BibitemOpen
  \bibfield  {author} {\bibinfo {author} {\bibfnamefont {C.-H.}\ \bibnamefont
  {Toh}}, \bibinfo {author} {\bibfnamefont {H.}~\bibnamefont {Zhang}}, \bibinfo
  {author} {\bibfnamefont {J.}~\bibnamefont {Lin}}, \bibinfo {author}
  {\bibfnamefont {A.~S.}\ \bibnamefont {Mayorov}}, \bibinfo {author}
  {\bibfnamefont {Y.-P.}\ \bibnamefont {Wang}}, \bibinfo {author}
  {\bibfnamefont {C.~M.}\ \bibnamefont {Orofeo}}, \bibinfo {author}
  {\bibfnamefont {D.~B.}\ \bibnamefont {Ferry}}, \bibinfo {author}
  {\bibfnamefont {H.}~\bibnamefont {Andersen}}, \bibinfo {author}
  {\bibfnamefont {N.}~\bibnamefont {Kakenov}}, \bibinfo {author} {\bibfnamefont
  {Z.}~\bibnamefont {Guo}}, \bibinfo {author} {\bibfnamefont {I.~H.}\
  \bibnamefont {Abidi}}, \bibinfo {author} {\bibfnamefont {H.}~\bibnamefont
  {Sims}}, \bibinfo {author} {\bibfnamefont {K.}~\bibnamefont {Suenaga}},
  \bibinfo {author} {\bibfnamefont {S.~T.}\ \bibnamefont {Pantelides}},\ and\
  \bibinfo {author} {\bibnamefont {Özyilmaz B.}},\ }\href
  {https://doi.org/10.1038/s41586-019-1871-2} {\bibfield  {journal} {\bibinfo
  {journal} {Nature}\ }\textbf {\bibinfo {volume} {577}},\ \bibinfo {pages}
  {199} (\bibinfo {year} {2020})}\BibitemShut {NoStop}%
\bibitem [{\citenamefont {Huang}\ \emph {et~al.}(2019)\citenamefont {Huang},
  \citenamefont {Zhang}, \citenamefont {Liu}, \citenamefont {Zhang},
  \citenamefont {Jin}, \citenamefont {Wang}, \citenamefont {Jiang},
  \citenamefont {Fan},\ and\ \citenamefont {Li}}]{2DA-Ex-Am-MoS2}%
  \BibitemOpen
  \bibfield  {author} {\bibinfo {author} {\bibfnamefont {Z.}~\bibnamefont
  {Huang}}, \bibinfo {author} {\bibfnamefont {T.}~\bibnamefont {Zhang}},
  \bibinfo {author} {\bibfnamefont {J.}~\bibnamefont {Liu}}, \bibinfo {author}
  {\bibfnamefont {L.}~\bibnamefont {Zhang}}, \bibinfo {author} {\bibfnamefont
  {Y.}~\bibnamefont {Jin}}, \bibinfo {author} {\bibfnamefont {J.}~\bibnamefont
  {Wang}}, \bibinfo {author} {\bibfnamefont {K.}~\bibnamefont {Jiang}},
  \bibinfo {author} {\bibfnamefont {S.}~\bibnamefont {Fan}},\ and\ \bibinfo
  {author} {\bibfnamefont {Q.}~\bibnamefont {Li}},\ }\href
  {https://doi.org/10.1021/acsaelm.9b00247} {\bibfield  {journal} {\bibinfo
  {journal} {Appl. Electron. Mater.}\ }\textbf {\bibinfo {volume} {1}},\
  \bibinfo {pages} {1314} (\bibinfo {year} {2019})}\BibitemShut {NoStop}%
\bibitem [{\citenamefont {Yang}\ \emph {et~al.}(2015)\citenamefont {Yang},
  \citenamefont {Hao}, \citenamefont {Yuan}, \citenamefont {Lin}, \citenamefont
  {Yau}, \citenamefont {Dai},\ and\ \citenamefont
  {Lau}}]{2DA-Ex-Am-B.phosphor}%
  \BibitemOpen
  \bibfield  {author} {\bibinfo {author} {\bibfnamefont {Z.}~\bibnamefont
  {Yang}}, \bibinfo {author} {\bibfnamefont {J.}~\bibnamefont {Hao}}, \bibinfo
  {author} {\bibfnamefont {S.}~\bibnamefont {Yuan}}, \bibinfo {author}
  {\bibfnamefont {S.}~\bibnamefont {Lin}}, \bibinfo {author} {\bibfnamefont
  {H.~M.}\ \bibnamefont {Yau}}, \bibinfo {author} {\bibfnamefont
  {J.}~\bibnamefont {Dai}},\ and\ \bibinfo {author} {\bibfnamefont {S.~P.}\
  \bibnamefont {Lau}},\ }\href {https://doi.org/10.1002/adma.201500990}
  {\bibfield  {journal} {\bibinfo  {journal} {Adv. Mater.}\ }\textbf {\bibinfo
  {volume} {27}},\ \bibinfo {pages} {3748} (\bibinfo {year}
  {2015})}\BibitemShut {NoStop}%
\bibitem [{\citenamefont {Chattopadhyay}\ \emph {et~al.}(2014)\citenamefont
  {Chattopadhyay}, \citenamefont {Banerjee}, \citenamefont {Das},\ and\
  \citenamefont {Sarkara}}]{2DA-Ex-Am-AgraCathod}%
  \BibitemOpen
  \bibfield  {author} {\bibinfo {author} {\bibfnamefont {K.~K.}\ \bibnamefont
  {Chattopadhyay}}, \bibinfo {author} {\bibfnamefont {D.}~\bibnamefont
  {Banerjee}}, \bibinfo {author} {\bibfnamefont {N.~S.}\ \bibnamefont {Das}},\
  and\ \bibinfo {author} {\bibfnamefont {D.}~\bibnamefont {Sarkara}},\ }\href
  {https://doi.org/10.1016/j.carbon.2013.12.082} {\bibfield  {journal}
  {\bibinfo  {journal} {Carbon}\ }\textbf {\bibinfo {volume} {72}},\ \bibinfo
  {pages} {4} (\bibinfo {year} {2014})}\BibitemShut {NoStop}%
\bibitem [{\citenamefont {Bhunia}\ \emph {et~al.}(2018)\citenamefont {Bhunia},
  \citenamefont {Panigrahi}, \citenamefont {Das}, \citenamefont
  {Chattopadhyay},\ and\ \citenamefont {Chattopadhyay}}]{2DA-Ex-Am-AgraOil}%
  \BibitemOpen
  \bibfield  {author} {\bibinfo {author} {\bibfnamefont {M.~M.}\ \bibnamefont
  {Bhunia}}, \bibinfo {author} {\bibfnamefont {K.}~\bibnamefont {Panigrahi}},
  \bibinfo {author} {\bibfnamefont {S.}~\bibnamefont {Das}}, \bibinfo {author}
  {\bibfnamefont {K.~K.}\ \bibnamefont {Chattopadhyay}},\ and\ \bibinfo
  {author} {\bibfnamefont {P.}~\bibnamefont {Chattopadhyay}},\ }\href
  {https://doi.org/10.1016/j.carbon.2018.08.012} {\bibfield  {journal}
  {\bibinfo  {journal} {Carbon}\ }\textbf {\bibinfo {volume} {139}},\ \bibinfo
  {pages} {1010} (\bibinfo {year} {2018})}\BibitemShut {NoStop}%
\bibitem [{\citenamefont {Wu}\ \emph {et~al.}(2019)\citenamefont {Wu},
  \citenamefont {Longo}, \citenamefont {Dzade}, \citenamefont {Sharma},
  \citenamefont {Hendrix}, \citenamefont {Bol}, \citenamefont {de~Leeuw},
  \citenamefont {Hensen},\ and\ \citenamefont
  {Hofmann}}]{2DA-Ex-Am-MoS2-hydrogen}%
  \BibitemOpen
  \bibfield  {author} {\bibinfo {author} {\bibfnamefont {L.}~\bibnamefont
  {Wu}}, \bibinfo {author} {\bibfnamefont {A.}~\bibnamefont {Longo}}, \bibinfo
  {author} {\bibfnamefont {N.~Y.}\ \bibnamefont {Dzade}}, \bibinfo {author}
  {\bibfnamefont {A.}~\bibnamefont {Sharma}}, \bibinfo {author} {\bibfnamefont
  {M.~M. R.~M.}\ \bibnamefont {Hendrix}}, \bibinfo {author} {\bibfnamefont
  {A.~A.}\ \bibnamefont {Bol}}, \bibinfo {author} {\bibfnamefont {N.~H.}\
  \bibnamefont {de~Leeuw}}, \bibinfo {author} {\bibfnamefont {E.~J.~M.}\
  \bibnamefont {Hensen}},\ and\ \bibinfo {author} {\bibfnamefont {J.~P.}\
  \bibnamefont {Hofmann}},\ }\href {https://doi.org/10.1002/cssc.201901811}
  {\bibfield  {journal} {\bibinfo  {journal} {Chem. Sus. Chem.}\ }\textbf
  {\bibinfo {volume} {12}},\ \bibinfo {pages} {4383} (\bibinfo {year}
  {2019})}\BibitemShut {NoStop}%
\bibitem [{\citenamefont {Fu}\ \emph {et~al.}(2019)\citenamefont {Fu},
  \citenamefont {Yang}, \citenamefont {Yang}, \citenamefont {Guo},\ and\
  \citenamefont {Huang}}]{2DA-Ex-Am-MoS3-toxic}%
  \BibitemOpen
  \bibfield  {author} {\bibinfo {author} {\bibfnamefont {W.}~\bibnamefont
  {Fu}}, \bibinfo {author} {\bibfnamefont {S.}~\bibnamefont {Yang}}, \bibinfo
  {author} {\bibfnamefont {H.}~\bibnamefont {Yang}}, \bibinfo {author}
  {\bibfnamefont {B.}~\bibnamefont {Guo}},\ and\ \bibinfo {author}
  {\bibfnamefont {Z.}~\bibnamefont {Huang}},\ }\href
  {https://doi.org/10.1039/C9TA05861C} {\bibfield  {journal} {\bibinfo
  {journal} {J. Mater. Chem. A}\ }\textbf {\bibinfo {volume} {7}},\ \bibinfo
  {pages} {18799} (\bibinfo {year} {2019})}\BibitemShut {NoStop}%
\bibitem [{\citenamefont {Glavin}\ \emph {et~al.}(2016)\citenamefont {Glavin},
  \citenamefont {Muratore}, \citenamefont {Jespersen}, \citenamefont {Hu},
  \citenamefont {Hagerty}, \citenamefont {Hilton},\ and\ \citenamefont
  {Blake}}]{2DA-Ex-Am-BN-Nano}%
  \BibitemOpen
  \bibfield  {author} {\bibinfo {author} {\bibfnamefont {N.~R.}\ \bibnamefont
  {Glavin}}, \bibinfo {author} {\bibfnamefont {C.}~\bibnamefont {Muratore}},
  \bibinfo {author} {\bibfnamefont {M.~L.}\ \bibnamefont {Jespersen}}, \bibinfo
  {author} {\bibfnamefont {J.}~\bibnamefont {Hu}}, \bibinfo {author}
  {\bibfnamefont {P.~T.}\ \bibnamefont {Hagerty}}, \bibinfo {author}
  {\bibfnamefont {A.~M.}\ \bibnamefont {Hilton}},\ and\ \bibinfo {author}
  {\bibfnamefont {A.~T.}\ \bibnamefont {Blake}},\ }\href
  {https://doi.org/10.1002/adfm.201505455} {\bibfield  {journal} {\bibinfo
  {journal} {Adv. Func. Mater.}\ }\textbf {\bibinfo {volume} {26}},\ \bibinfo
  {pages} {2640} (\bibinfo {year} {2016})}\BibitemShut {NoStop}%
\bibitem [{\citenamefont {Antidormi}\ and\ \citenamefont
  {Roche}(2021)}]{2DM-Agra}%
  \BibitemOpen
  \bibfield  {author} {\bibinfo {author} {\bibfnamefont {L.}~\bibnamefont
  {Antidormi}, \bibfnamefont {A.~Colombo}}\ and\ \bibinfo {author}
  {\bibfnamefont {S.}~\bibnamefont {Roche}},\ }\href
  {https://doi.org/10.1088/2053-1583/abc7f8} {\bibfield  {journal} {\bibinfo
  {journal} {2D Mater.}\ }\textbf {\bibinfo {volume} {8}},\ \bibinfo {pages}
  {015028} (\bibinfo {year} {2021})}\BibitemShut {NoStop}%
\bibitem [{\citenamefont {Zhu}\ and\ \citenamefont
  {Ertekin}(2016)}]{Nano-Agra}%
  \BibitemOpen
  \bibfield  {author} {\bibinfo {author} {\bibfnamefont {T.}~\bibnamefont
  {Zhu}}\ and\ \bibinfo {author} {\bibfnamefont {E.}~\bibnamefont {Ertekin}},\
  }\href {https://doi.org/10.1021/acs.nanolett.6b00557} {\bibfield  {journal}
  {\bibinfo  {journal} {Nano Lett.}\ }\textbf {\bibinfo {volume} {16}},\
  \bibinfo {pages} {4763} (\bibinfo {year} {2016})}\BibitemShut {NoStop}%
\bibitem [{\citenamefont {Mortazavi}\ \emph {et~al.}(2016)\citenamefont
  {Mortazavi}, \citenamefont {Z.}, \citenamefont {C.~Pereira}, \citenamefont
  {Harju},\ and\ \citenamefont {Rabczuk}}]{Carbonmech-Agra}%
  \BibitemOpen
  \bibfield  {author} {\bibinfo {author} {\bibfnamefont {B.}~\bibnamefont
  {Mortazavi}}, \bibinfo {author} {\bibfnamefont {F.}~\bibnamefont {Z.}},
  \bibinfo {author} {\bibfnamefont {L.~F.}\ \bibnamefont {C.~Pereira}},
  \bibinfo {author} {\bibfnamefont {A.}~\bibnamefont {Harju}},\ and\ \bibinfo
  {author} {\bibfnamefont {T.}~\bibnamefont {Rabczuk}},\ }\href
  {https://doi.org/10.1016/j.carbon.2016.03.007} {\bibfield  {journal}
  {\bibinfo  {journal} {Carbon}\ }\textbf {\bibinfo {volume} {103}},\ \bibinfo
  {pages} {318–326} (\bibinfo {year} {2016})}\BibitemShut {NoStop}%
\bibitem [{\citenamefont {Bhattarai}\ \emph {et~al.}(2018)\citenamefont
  {Bhattarai}, \citenamefont {Biswas}, \citenamefont {Atta-Fynn},\ and\
  \citenamefont {Drabold}}]{2DM-2El}%
  \BibitemOpen
  \bibfield  {author} {\bibinfo {author} {\bibfnamefont {B.}~\bibnamefont
  {Bhattarai}}, \bibinfo {author} {\bibfnamefont {P.}~\bibnamefont {Biswas}},
  \bibinfo {author} {\bibfnamefont {R.}~\bibnamefont {Atta-Fynn}},\ and\
  \bibinfo {author} {\bibfnamefont {D.~A.}\ \bibnamefont {Drabold}},\ }\href
  {https://doi.org/10.1039/c8cp02545b} {\bibfield  {journal} {\bibinfo
  {journal} {Phys. Chem. Chem. Phys.}\ }\textbf {\bibinfo {volume} {20}},\
  \bibinfo {pages} {19546} (\bibinfo {year} {2018})}\BibitemShut {NoStop}%
\bibitem [{\citenamefont {Ravinder}\ \emph {et~al.}(2019)\citenamefont
  {Ravinder}, \citenamefont {Kumar}, \citenamefont {M.},\ and\ \citenamefont
  {Krishnan}}]{2d-2019-Agra}%
  \BibitemOpen
  \bibfield  {author} {\bibinfo {author} {\bibfnamefont {R.}~\bibnamefont
  {Ravinder}}, \bibinfo {author} {\bibfnamefont {R.}~\bibnamefont {Kumar}},
  \bibinfo {author} {\bibfnamefont {A.}~\bibnamefont {M.}},\ and\ \bibinfo
  {author} {\bibfnamefont {N.~M.~A.}\ \bibnamefont {Krishnan}},\ }\href
  {https://doi.org/10.1038/s41598-019-41231-z} {\bibfield  {journal} {\bibinfo
  {journal} {Sci. Rep.}\ }\textbf {\bibinfo {volume} {9}},\ \bibinfo {pages}
  {4517} (\bibinfo {year} {2019})}\BibitemShut {NoStop}%
\bibitem [{\citenamefont {Geder}\ \emph {et~al.}(2014)\citenamefont {Geder},
  \citenamefont {Kotakoski}, \citenamefont {Kaiser},\ and\ \citenamefont
  {Meyer}}]{CGlass-struct}%
  \BibitemOpen
  \bibfield  {author} {\bibinfo {author} {\bibfnamefont {F.~R.}\ \bibnamefont
  {Geder}}, \bibinfo {author} {\bibfnamefont {J.}~\bibnamefont {Kotakoski}},
  \bibinfo {author} {\bibfnamefont {U.}~\bibnamefont {Kaiser}},\ and\ \bibinfo
  {author} {\bibfnamefont {J.~C.}\ \bibnamefont {Meyer}},\ }\href
  {https://doi.org/10.1038/srep04060} {\bibfield  {journal} {\bibinfo
  {journal} {Sci. Rep.}\ }\textbf {\bibinfo {volume} {4}},\ \bibinfo {pages}
  {4060} (\bibinfo {year} {2014})}\BibitemShut {NoStop}%
\bibitem [{\citenamefont {Hoang}\ and\ \citenamefont {Long}(2016)}]{Si1}%
  \BibitemOpen
  \bibfield  {author} {\bibinfo {author} {\bibfnamefont {V.~V.}\ \bibnamefont
  {Hoang}}\ and\ \bibinfo {author} {\bibfnamefont {N.~T.}\ \bibnamefont
  {Long}},\ }\href {https://doi.org/10.1088/0953-8984/28/19/195401} {\bibfield
  {journal} {\bibinfo  {journal} {J. Phys.: Condens. Matter}\ }\textbf
  {\bibinfo {volume} {28}},\ \bibinfo {pages} {195401} (\bibinfo {year}
  {2016})}\BibitemShut {NoStop}%
\bibitem [{\citenamefont {Gao}\ \emph {et~al.}(2018)\citenamefont {Gao},
  \citenamefont {Zhou}, \citenamefont {Zhang},\ and\ \citenamefont {Hu}}]{Si2}%
  \BibitemOpen
  \bibfield  {author} {\bibinfo {author} {\bibfnamefont {Y.}~\bibnamefont
  {Gao}}, \bibinfo {author} {\bibfnamefont {Y.}~\bibnamefont {Zhou}}, \bibinfo
  {author} {\bibfnamefont {X.}~\bibnamefont {Zhang}},\ and\ \bibinfo {author}
  {\bibfnamefont {M.}~\bibnamefont {Hu}},\ }\href
  {https://doi.org/10.1021/acs.jpcc.8b01466} {\bibfield  {journal} {\bibinfo
  {journal} {J. Phys. Chem. C}\ }\textbf {\bibinfo {volume} {122}},\ \bibinfo
  {pages} {9220−9228} (\bibinfo {year} {2018})}\BibitemShut {NoStop}%
\bibitem [{\citenamefont {Zhou}\ and\ \citenamefont {Hu}(2016)}]{Si3}%
  \BibitemOpen
  \bibfield  {author} {\bibinfo {author} {\bibfnamefont {Y.}~\bibnamefont
  {Zhou}}\ and\ \bibinfo {author} {\bibfnamefont {M.}~\bibnamefont {Hu}},\
  }\href {https://doi.org/10.1021/acs.nanolett.6b02450} {\bibfield  {journal}
  {\bibinfo  {journal} {Nano Lett.}\ }\textbf {\bibinfo {volume} {16}},\
  \bibinfo {pages} {6178−6187} (\bibinfo {year} {2016})}\BibitemShut
  {NoStop}%
\bibitem [{\citenamefont {Hoang}\ \emph {et~al.}(2020)\citenamefont {Hoang},
  \citenamefont {Giang},\ and\ \citenamefont {Dong}}]{Si-penta}%
  \BibitemOpen
  \bibfield  {author} {\bibinfo {author} {\bibfnamefont {V.~V.}\ \bibnamefont
  {Hoang}}, \bibinfo {author} {\bibfnamefont {N.~H.}\ \bibnamefont {Giang}},\
  and\ \bibinfo {author} {\bibfnamefont {T.~Q.}\ \bibnamefont {Dong}},\ }\href
  {https://doi.org/10.1080/14786435.2020.1750724} {\bibfield  {journal}
  {\bibinfo  {journal} {Philos. Mag.}\ }\textbf {\bibinfo {volume} {100}},\
  \bibinfo {pages} {1962} (\bibinfo {year} {2020})}\BibitemShut {NoStop}%
\bibitem [{\citenamefont {Long}\ \emph {et~al.}(2018)\citenamefont {Long},
  \citenamefont {Huy}, \citenamefont {Tuan}, \citenamefont {Le}, \citenamefont
  {Hoang},\ and\ \citenamefont {H.}}]{Si-structure}%
  \BibitemOpen
  \bibfield  {author} {\bibinfo {author} {\bibfnamefont {N.~T.}\ \bibnamefont
  {Long}}, \bibinfo {author} {\bibfnamefont {H.~A.}\ \bibnamefont {Huy}},
  \bibinfo {author} {\bibfnamefont {T.~Q.}\ \bibnamefont {Tuan}}, \bibinfo
  {author} {\bibfnamefont {O.~K.}\ \bibnamefont {Le}}, \bibinfo {author}
  {\bibfnamefont {V.~V.}\ \bibnamefont {Hoang}},\ and\ \bibinfo {author}
  {\bibfnamefont {G.~N.}\ \bibnamefont {H.}},\ }\href
  {https://doi.org/10.1016/j.jnoncrysol.2018.02.024} {\bibfield  {journal}
  {\bibinfo  {journal} {J. Non-Cryst. Solids}\ }\textbf {\bibinfo {volume}
  {487}},\ \bibinfo {pages} {87} (\bibinfo {year} {2018})}\BibitemShut
  {NoStop}%
\bibitem [{\citenamefont {Huy}\ \emph {et~al.}(2019)\citenamefont {Huy},
  \citenamefont {Nguyen}, \citenamefont {Nguyen}, \citenamefont {Truong},
  \citenamefont {Ong}, \citenamefont {Hoang},\ and\ \citenamefont
  {H.}}]{Si-structure2}%
  \BibitemOpen
  \bibfield  {author} {\bibinfo {author} {\bibfnamefont {H.~A.}\ \bibnamefont
  {Huy}}, \bibinfo {author} {\bibfnamefont {L.~T.}\ \bibnamefont {Nguyen}},
  \bibinfo {author} {\bibfnamefont {D.~L.~T.}\ \bibnamefont {Nguyen}}, \bibinfo
  {author} {\bibfnamefont {T.~Q.}\ \bibnamefont {Truong}}, \bibinfo {author}
  {\bibfnamefont {L.~K.}\ \bibnamefont {Ong}}, \bibinfo {author} {\bibfnamefont
  {V.~V.}\ \bibnamefont {Hoang}},\ and\ \bibinfo {author} {\bibfnamefont
  {G.~N.}\ \bibnamefont {H.}},\ }\href
  {https://doi.org/10.1088/1361-648X/aaf402} {\bibfield  {journal} {\bibinfo
  {journal} {J. Phys.: Condens. Matter.}\ }\textbf {\bibinfo {volume} {31}},\
  \bibinfo {pages} {095403} (\bibinfo {year} {2019})}\BibitemShut {NoStop}%
\bibitem [{\citenamefont {Li}\ and\ \citenamefont {Yue}(2014)}]{SiC-thermal}%
  \BibitemOpen
  \bibfield  {author} {\bibinfo {author} {\bibfnamefont {M.}~\bibnamefont
  {Li}}\ and\ \bibinfo {author} {\bibfnamefont {Y.}~\bibnamefont {Yue}},\
  }\href {https://doi.org/10.1039/c4ra02985b} {\bibfield  {journal} {\bibinfo
  {journal} {RSC Adv.}\ }\textbf {\bibinfo {volume} {4}},\ \bibinfo {pages}
  {23010–23016} (\bibinfo {year} {2014})}\BibitemShut {NoStop}%
\bibitem [{\citenamefont {Tranh}\ \emph {et~al.}(2021)\citenamefont {Tranh},
  \citenamefont {Hoang},\ and\ \citenamefont {Hanh}}]{SiC-nanoribbon1}%
  \BibitemOpen
  \bibfield  {author} {\bibinfo {author} {\bibfnamefont {D.~T.~N.}\
  \bibnamefont {Tranh}}, \bibinfo {author} {\bibfnamefont {V.~V.}\ \bibnamefont
  {Hoang}},\ and\ \bibinfo {author} {\bibfnamefont {T.~T.~T.}\ \bibnamefont
  {Hanh}},\ }\href {https://doi.org/10.1016/j.physb.2020.412746} {\bibfield
  {journal} {\bibinfo  {journal} {Phys. B: Condens. Matter}\ }\textbf {\bibinfo
  {volume} {608}},\ \bibinfo {pages} {412746} (\bibinfo {year}
  {2021})}\BibitemShut {NoStop}%
\bibitem [{\citenamefont {Hoang}\ \emph {et~al.}(2022)\citenamefont {Hoang},
  \citenamefont {Giang}, \citenamefont {Dong},\ and\ \citenamefont
  {Bubanja}}]{SiC-nanoribbon2}%
  \BibitemOpen
  \bibfield  {author} {\bibinfo {author} {\bibfnamefont {V.~V.}\ \bibnamefont
  {Hoang}}, \bibinfo {author} {\bibfnamefont {N.~H.}\ \bibnamefont {Giang}},
  \bibinfo {author} {\bibfnamefont {T.~Q.}\ \bibnamefont {Dong}},\ and\
  \bibinfo {author} {\bibfnamefont {V.}~\bibnamefont {Bubanja}},\ }\href
  {https://doi.org/10.1016/j.commatsci.2021.111123} {\bibfield  {journal}
  {\bibinfo  {journal} {Comput. Mater. Sci.}\ }\textbf {\bibinfo {volume}
  {203}},\ \bibinfo {pages} {111123} (\bibinfo {year} {2022})}\BibitemShut
  {NoStop}%
\bibitem [{\citenamefont {Giang}\ and\ \citenamefont
  {Hoang}(2021)}]{2DA-Germanene}%
  \BibitemOpen
  \bibfield  {author} {\bibinfo {author} {\bibfnamefont {N.~H.}\ \bibnamefont
  {Giang}}\ and\ \bibinfo {author} {\bibfnamefont {V.~V.}\ \bibnamefont
  {Hoang}},\ }\href {https://doi.org/10.1016/j.physe.2020.114492} {\bibfield
  {journal} {\bibinfo  {journal} {Physica E Low Dimens. Syst. Nanostruct.}\
  }\textbf {\bibinfo {volume} {126}},\ \bibinfo {pages} {114492} (\bibinfo
  {year} {2021})}\BibitemShut {NoStop}%
\bibitem [{\citenamefont {Durandurdu}(2015)}]{2DA-Ex-Am-BN}%
  \BibitemOpen
  \bibfield  {author} {\bibinfo {author} {\bibfnamefont {M.}~\bibnamefont
  {Durandurdu}},\ }\href {https://doi.org/10.1016/j.jnoncrysol.2015.07.033}
  {\bibfield  {journal} {\bibinfo  {journal} {J. Non-Cryst. Solids}\ }\textbf
  {\bibinfo {volume} {427}},\ \bibinfo {pages} {41} (\bibinfo {year}
  {2015})}\BibitemShut {NoStop}%
\bibitem [{\citenamefont {Jana}\ \emph {et~al.}(2019)\citenamefont {Jana},
  \citenamefont {Savio}, \citenamefont {Deringer},\ and\ \citenamefont
  {Pastewka}}]{3D-Mech-Agra}%
  \BibitemOpen
  \bibfield  {author} {\bibinfo {author} {\bibfnamefont {R.}~\bibnamefont
  {Jana}}, \bibinfo {author} {\bibfnamefont {D.}~\bibnamefont {Savio}},
  \bibinfo {author} {\bibfnamefont {V.~L.}\ \bibnamefont {Deringer}},\ and\
  \bibinfo {author} {\bibfnamefont {L.}~\bibnamefont {Pastewka}},\ }\href
  {https://doi.org/10.1088/1361-651X/ab45da} {\bibfield  {journal} {\bibinfo
  {journal} {Modelling Simul. Mater. Sci. Eng.}\ }\textbf {\bibinfo {volume}
  {27}},\ \bibinfo {pages} {085009} (\bibinfo {year} {2019})}\BibitemShut
  {NoStop}%
\bibitem [{\citenamefont {Plimpton}(1995)}]{lammps}%
  \BibitemOpen
  \bibfield  {author} {\bibinfo {author} {\bibfnamefont {S.}~\bibnamefont
  {Plimpton}},\ }\href {https://doi.org/10.1006/jcph.1995.1039} {\bibfield
  {journal} {\bibinfo  {journal} {J. Comput. Phys.}\ }\textbf {\bibinfo
  {volume} {117}},\ \bibinfo {pages} {1} (\bibinfo {year} {1995})}\BibitemShut
  {NoStop}%
\bibitem [{\citenamefont {Tersoff}(1994)}]{tersoff}%
  \BibitemOpen
  \bibfield  {author} {\bibinfo {author} {\bibfnamefont {J.}~\bibnamefont
  {Tersoff}},\ }\href {https://doi.org/10.1103/PhysRevB.49.16349} {\bibfield
  {journal} {\bibinfo  {journal} {Phys. Rev. B}\ }\textbf {\bibinfo {volume}
  {49}},\ \bibinfo {pages} {16349–16352} (\bibinfo {year}
  {1994})}\BibitemShut {NoStop}%
\bibitem [{\citenamefont {Berendsen}\ \emph {et~al.}(1984)\citenamefont
  {Berendsen}, \citenamefont {Postma}, \citenamefont {van Gunsteren},
  \citenamefont {Dinola},\ and\ \citenamefont {Haak}}]{berendsen}%
  \BibitemOpen
  \bibfield  {author} {\bibinfo {author} {\bibfnamefont {H.~J.~C.}\
  \bibnamefont {Berendsen}}, \bibinfo {author} {\bibfnamefont {J.~P.~M.}\
  \bibnamefont {Postma}}, \bibinfo {author} {\bibfnamefont {W.}~\bibnamefont
  {van Gunsteren}}, \bibinfo {author} {\bibfnamefont {A.}~\bibnamefont
  {Dinola}},\ and\ \bibinfo {author} {\bibfnamefont {J.~R.}\ \bibnamefont
  {Haak}},\ }\href {https://doi.org/10.1063/1.448118} {\bibfield  {journal}
  {\bibinfo  {journal} {J. Chem. Phys.}\ }\textbf {\bibinfo {volume} {81}},\
  \bibinfo {pages} {3684} (\bibinfo {year} {1984})}\BibitemShut {NoStop}%
\bibitem [{\citenamefont {Kresse}\ and\ \citenamefont {Hafner}(1994)}]{vasp2}%
  \BibitemOpen
  \bibfield  {author} {\bibinfo {author} {\bibfnamefont {G.}~\bibnamefont
  {Kresse}}\ and\ \bibinfo {author} {\bibfnamefont {J.}~\bibnamefont
  {Hafner}},\ }\href {https://doi.org/10.1103/PhysRevB.49.14251} {\bibfield
  {journal} {\bibinfo  {journal} {Phys. Rev. B}\ }\textbf {\bibinfo {volume}
  {49}},\ \bibinfo {pages} {14251} (\bibinfo {year} {1994})}\BibitemShut
  {NoStop}%
\bibitem [{\citenamefont {Kresse}\ and\ \citenamefont
  {Furthm\"~uller}(1996)}]{vasp3}%
  \BibitemOpen
  \bibfield  {author} {\bibinfo {author} {\bibfnamefont {G.}~\bibnamefont
  {Kresse}}\ and\ \bibinfo {author} {\bibfnamefont {J.}~\bibnamefont
  {Furthm\"~uller}},\ }\href {https://doi.org/10.1016/0927-0256(96)00008-0}
  {\bibfield  {journal} {\bibinfo  {journal} {Comput. Mater. Sci.}\ }\textbf
  {\bibinfo {volume} {6}},\ \bibinfo {pages} {15} (\bibinfo {year}
  {1996})}\BibitemShut {NoStop}%
\bibitem [{\citenamefont {Bl\"ochl}(1994)}]{blochl}%
  \BibitemOpen
  \bibfield  {author} {\bibinfo {author} {\bibfnamefont {P.~E.}\ \bibnamefont
  {Bl\"ochl}},\ }\href {https://doi.org/10.1103/PhysRevB.50.17953} {\bibfield
  {journal} {\bibinfo  {journal} {Phys. Rev. B}\ }\textbf {\bibinfo {volume}
  {50}},\ \bibinfo {pages} {17953} (\bibinfo {year} {1994})}\BibitemShut
  {NoStop}%
\bibitem [{\citenamefont {Perdew}\ \emph
  {et~al.}(1996{\natexlab{a}})\citenamefont {Perdew}, \citenamefont {Burke},\
  and\ \citenamefont {Ernzerhof}}]{pbe}%
  \BibitemOpen
  \bibfield  {author} {\bibinfo {author} {\bibfnamefont {J.~P.}\ \bibnamefont
  {Perdew}}, \bibinfo {author} {\bibfnamefont {K.}~\bibnamefont {Burke}},\ and\
  \bibinfo {author} {\bibfnamefont {M.}~\bibnamefont {Ernzerhof}},\ }\href
  {https://doi.org/10.1103/PhysRevLett.77.3865} {\bibfield  {journal} {\bibinfo
   {journal} {Phys. Rev. Lett.}\ }\textbf {\bibinfo {volume} {77}},\ \bibinfo
  {pages} {3865} (\bibinfo {year} {1996}{\natexlab{a}})}\BibitemShut {NoStop}%
\bibitem [{\citenamefont {Perdew}\ \emph
  {et~al.}(1996{\natexlab{b}})\citenamefont {Perdew}, \citenamefont {Burke},\
  and\ \citenamefont {Ernzerhof}}]{monk}%
  \BibitemOpen
  \bibfield  {author} {\bibinfo {author} {\bibfnamefont {J.~P.}\ \bibnamefont
  {Perdew}}, \bibinfo {author} {\bibfnamefont {K.}~\bibnamefont {Burke}},\ and\
  \bibinfo {author} {\bibfnamefont {M.}~\bibnamefont {Ernzerhof}},\ }\href
  {https://doi.org/10.1103/PhysRevLett.77.3865} {\bibfield  {journal} {\bibinfo
   {journal} {Phys. Rev. Lett.}\ }\textbf {\bibinfo {volume} {77}},\ \bibinfo
  {pages} {3865} (\bibinfo {year} {1996}{\natexlab{b}})}\BibitemShut {NoStop}%
\bibitem [{\citenamefont {Chen}\ \emph {et~al.}(2013)\citenamefont {Chen},
  \citenamefont {Tian}, \citenamefont {Persson}, \citenamefont {Duan},\ and\
  \citenamefont {Chen}}]{bindingenergy}%
  \BibitemOpen
  \bibfield  {author} {\bibinfo {author} {\bibfnamefont {X.}~\bibnamefont
  {Chen}}, \bibinfo {author} {\bibfnamefont {F.}~\bibnamefont {Tian}}, \bibinfo
  {author} {\bibfnamefont {C.}~\bibnamefont {Persson}}, \bibinfo {author}
  {\bibfnamefont {W.}~\bibnamefont {Duan}},\ and\ \bibinfo {author}
  {\bibfnamefont {N.-X.}\ \bibnamefont {Chen}},\ }\href
  {https://doi.org/10.1038/srep03046} {\bibfield  {journal} {\bibinfo
  {journal} {Sci. Rep.}\ }\textbf {\bibinfo {volume} {3}},\ \bibinfo {pages}
  {3046} (\bibinfo {year} {2013})}\BibitemShut {NoStop}%
\bibitem [{\citenamefont {Gajdo\ifmmode~\check{s}\else \v{s}\fi{}}\ \emph
  {et~al.}(2006)\citenamefont {Gajdo\ifmmode~\check{s}\else \v{s}\fi{}},
  \citenamefont {Hummer}, \citenamefont {Kresse}, \citenamefont
  {Furthm\"uller},\ and\ \citenamefont {Bechstedt}}]{optic1}%
  \BibitemOpen
  \bibfield  {author} {\bibinfo {author} {\bibfnamefont {M.}~\bibnamefont
  {Gajdo\ifmmode~\check{s}\else \v{s}\fi{}}}, \bibinfo {author} {\bibfnamefont
  {K.}~\bibnamefont {Hummer}}, \bibinfo {author} {\bibfnamefont
  {G.}~\bibnamefont {Kresse}}, \bibinfo {author} {\bibfnamefont
  {J.}~\bibnamefont {Furthm\"uller}},\ and\ \bibinfo {author} {\bibfnamefont
  {F.}~\bibnamefont {Bechstedt}},\ }\href
  {https://doi.org/10.1103/PhysRevB.73.045112} {\bibfield  {journal} {\bibinfo
  {journal} {Phys. Rev. B}\ }\textbf {\bibinfo {volume} {73}},\ \bibinfo
  {pages} {045112} (\bibinfo {year} {2006})}\BibitemShut {NoStop}%
\bibitem [{\citenamefont {Shahrokhi}\ and\ \citenamefont
  {Leonard}(2017)}]{Gra-SiC}%
  \BibitemOpen
  \bibfield  {author} {\bibinfo {author} {\bibfnamefont {M.}~\bibnamefont
  {Shahrokhi}}\ and\ \bibinfo {author} {\bibfnamefont {C.}~\bibnamefont
  {Leonard}},\ }\href {https://doi.org/10.1016/j.jallcom.2016.10.101}
  {\bibfield  {journal} {\bibinfo  {journal} {Journal of Alloys and Compounds}\
  }\textbf {\bibinfo {volume} {693}},\ \bibinfo {pages} {1185} (\bibinfo {year}
  {2017})}\BibitemShut {NoStop}%
\bibitem [{\citenamefont {Polley}\ \emph {et~al.}(2023)\citenamefont {Polley},
  \citenamefont {Fedderwitz}, \citenamefont {Balasubramanian}, \citenamefont
  {Zakharov}, \citenamefont {Yakimova}, \citenamefont {Bäcke}, \citenamefont
  {Ekman}, \citenamefont {Dash}, \citenamefont {Kubatkin},\ and\ \citenamefont
  {Lara-Avila}}]{SiC2023}%
  \BibitemOpen
  \bibfield  {author} {\bibinfo {author} {\bibfnamefont {M.}~\bibnamefont
  {Polley}}, \bibinfo {author} {\bibfnamefont {H.}~\bibnamefont {Fedderwitz}},
  \bibinfo {author} {\bibfnamefont {T.}~\bibnamefont {Balasubramanian}},
  \bibinfo {author} {\bibfnamefont {A.~A.}\ \bibnamefont {Zakharov}}, \bibinfo
  {author} {\bibfnamefont {R.}~\bibnamefont {Yakimova}}, \bibinfo {author}
  {\bibfnamefont {O.}~\bibnamefont {Bäcke}}, \bibinfo {author} {\bibfnamefont
  {J.}~\bibnamefont {Ekman}}, \bibinfo {author} {\bibfnamefont {S.~P.}\
  \bibnamefont {Dash}}, \bibinfo {author} {\bibfnamefont {S.}~\bibnamefont
  {Kubatkin}},\ and\ \bibinfo {author} {\bibfnamefont {S.}~\bibnamefont
  {Lara-Avila}},\ }\href {https://doi.org/10.1103/PhysRevLett.130.076203}
  {\bibfield  {journal} {\bibinfo  {journal} {Phys. Rev. Lett.}\ }\textbf
  {\bibinfo {volume} {130}},\ \bibinfo {pages} {076203} (\bibinfo {year}
  {2023})}\BibitemShut {NoStop}%
\bibitem [{\citenamefont {Mostaani}\ \emph {et~al.}(2015)\citenamefont
  {Mostaani}, \citenamefont {Drummond},\ and\ \citenamefont
  {I.}}]{bindingenergy2}%
  \BibitemOpen
  \bibfield  {author} {\bibinfo {author} {\bibfnamefont {E.}~\bibnamefont
  {Mostaani}}, \bibinfo {author} {\bibfnamefont {N.~D.}\ \bibnamefont
  {Drummond}},\ and\ \bibinfo {author} {\bibfnamefont {F.~V.}\ \bibnamefont
  {I.}},\ }\href {https://doi.org/10.1103/PhysRevLett.115.115501} {\bibfield
  {journal} {\bibinfo  {journal} {Phys. Rev. Lett.}\ }\textbf {\bibinfo
  {volume} {115}},\ \bibinfo {pages} {115501} (\bibinfo {year}
  {2015})}\BibitemShut {NoStop}%
\bibitem [{\citenamefont {Marinopoulos}\ \emph {et~al.}(2004)\citenamefont
  {Marinopoulos}, \citenamefont {Reining}, \citenamefont {Rubio},\ and\
  \citenamefont {Olevano}}]{Rubio}%
  \BibitemOpen
  \bibfield  {author} {\bibinfo {author} {\bibfnamefont {A.~G.}\ \bibnamefont
  {Marinopoulos}}, \bibinfo {author} {\bibfnamefont {L.}~\bibnamefont
  {Reining}}, \bibinfo {author} {\bibfnamefont {A.}~\bibnamefont {Rubio}},\
  and\ \bibinfo {author} {\bibfnamefont {V.}~\bibnamefont {Olevano}},\ }\href
  {https://doi.org/10.1103/PhysRevB.69.245419} {\bibfield  {journal} {\bibinfo
  {journal} {Phys. Rev. B}\ }\textbf {\bibinfo {volume} {69}},\ \bibinfo
  {pages} {245419} (\bibinfo {year} {2004})}\BibitemShut {NoStop}%
\bibitem [{\citenamefont {Tien}\ \emph {et~al.}(2022)\citenamefont {Tien},
  \citenamefont {Thao}, \citenamefont {Thuan},\ and\ \citenamefont
  {Chuong}}]{PentaGra-Optic}%
  \BibitemOpen
  \bibfield  {author} {\bibinfo {author} {\bibfnamefont {N.~T.}\ \bibnamefont
  {Tien}}, \bibinfo {author} {\bibfnamefont {P.~T.~B.}\ \bibnamefont {Thao}},
  \bibinfo {author} {\bibfnamefont {L.~V.~P.}\ \bibnamefont {Thuan}},\ and\
  \bibinfo {author} {\bibfnamefont {D.~H.}\ \bibnamefont {Chuong}},\ }\href
  {https://doi.org/10.1016/j.commatsci.2021.111065} {\bibfield  {journal}
  {\bibinfo  {journal} {Comput. Mater. Sci.}\ }\textbf {\bibinfo {volume}
  {15}},\ \bibinfo {pages} {111065} (\bibinfo {year} {2022})}\BibitemShut
  {NoStop}%
\bibitem [{\citenamefont {Ould~Ne}\ \emph {et~al.}(2017)\citenamefont
  {Ould~Ne}, \citenamefont {Abbasi}, \citenamefont {El~hachimi}, \citenamefont
  {Benyoussef}, \citenamefont {Ez-Zahraouy},\ and\ \citenamefont
  {El~Kenz}}]{Gra-Gedop}%
  \BibitemOpen
  \bibfield  {author} {\bibinfo {author} {\bibfnamefont {M.~L.}\ \bibnamefont
  {Ould~Ne}}, \bibinfo {author} {\bibfnamefont {A.}~\bibnamefont {Abbasi}},
  \bibinfo {author} {\bibfnamefont {A.}~\bibnamefont {El~hachimi}}, \bibinfo
  {author} {\bibfnamefont {A.}~\bibnamefont {Benyoussef}}, \bibinfo {author}
  {\bibfnamefont {H.}~\bibnamefont {Ez-Zahraouy}},\ and\ \bibinfo {author}
  {\bibfnamefont {A.}~\bibnamefont {El~Kenz}},\ }\href
  {https://doi.org/10.1007/s11082-017-1024-5} {\bibfield  {journal} {\bibinfo
  {journal} {Opt Quant Electron}\ }\textbf {\bibinfo {volume} {49}},\ \bibinfo
  {pages} {218} (\bibinfo {year} {2017})}\BibitemShut {NoStop}%
\bibitem [{\citenamefont {Mohan}\ \emph {et~al.}(2013)\citenamefont {Mohan},
  \citenamefont {Kumar},\ and\ \citenamefont {Ahluwalia}}]{Siopt}%
  \BibitemOpen
  \bibfield  {author} {\bibinfo {author} {\bibfnamefont {B.}~\bibnamefont
  {Mohan}}, \bibinfo {author} {\bibfnamefont {A.}~\bibnamefont {Kumar}},\ and\
  \bibinfo {author} {\bibfnamefont {P.}~\bibnamefont {Ahluwalia}},\ }\href
  {https://doi.org/10.1016/j.physe.2013.05.014} {\bibfield  {journal} {\bibinfo
   {journal} {Physica E}\ }\textbf {\bibinfo {volume} {53}},\ \bibinfo {pages}
  {233} (\bibinfo {year} {2013})}\BibitemShut {NoStop}%
\bibitem [{\citenamefont {Giri}\ \emph {et~al.}(2022)\citenamefont {Giri},
  \citenamefont {Dionne},\ and\ \citenamefont {Hopkins}}]{Giri-1}%
  \BibitemOpen
  \bibfield  {author} {\bibinfo {author} {\bibfnamefont {A.}~\bibnamefont
  {Giri}}, \bibinfo {author} {\bibfnamefont {C.~J.}\ \bibnamefont {Dionne}},\
  and\ \bibinfo {author} {\bibfnamefont {P.~E.}\ \bibnamefont {Hopkins}},\
  }\href {https://doi.org/10.1038/s41524-022-00741-7} {\bibfield  {journal}
  {\bibinfo  {journal} {Npj Comput. Mater.}\ }\textbf {\bibinfo {volume} {8}},\
  \bibinfo {pages} {55} (\bibinfo {year} {2022})}\BibitemShut {NoStop}%
\bibitem [{\citenamefont {Wingert}\ \emph {et~al.}(2016)\citenamefont
  {Wingert}, \citenamefont {Zheng}, \citenamefont {Kwon},\ and\ \citenamefont
  {Chen}}]{Si-SiC}%
  \BibitemOpen
  \bibfield  {author} {\bibinfo {author} {\bibfnamefont {M.~C.}\ \bibnamefont
  {Wingert}}, \bibinfo {author} {\bibfnamefont {J.}~\bibnamefont {Zheng}},
  \bibinfo {author} {\bibfnamefont {S.}~\bibnamefont {Kwon}},\ and\ \bibinfo
  {author} {\bibfnamefont {R.}~\bibnamefont {Chen}},\ }\href
  {https://doi.org/10.1088/0268-1242/31/11/113003} {\bibfield  {journal}
  {\bibinfo  {journal} {Semicond. Sci. Technol.}\ }\textbf {\bibinfo {volume}
  {31}},\ \bibinfo {pages} {113003} (\bibinfo {year} {2016})}\BibitemShut
  {NoStop}%
\end{thebibliography}%

%\section{Author contributions statement}

%\section{Additional information}

%\textbf{Competing interests} The authors declare no competing interests.

%\bibliography{autosam}           

% and a bib file to produce the 
                                 % bibliography (preferred). The
                                 % correct style is generated by
                                 % Elsevier at the time of printing.

%\begin{thebibliography}{99}     % Otherwise use the  
                                 % thebibliography environment.
                                 % Insert the full references here.
                                 % See a recent issue of Automatica 
                                 % for the style.
%  \bibitem[Heritage, 1992]{Heritage:92}
%     (1992) {\it The American Heritage. 
%     Dictionary of the American Language.}
%     Houghton Mifflin Company.
%  \bibitem[Able, 1956]{Abl:56}
%     B.~C.~Able (1956). Nucleic acid content of macroscope. 
%     {\it Nature 2}, 7--9. 
%  \bibitem[Able {\em et al.}, 1954]{AbTaRu:54}   
%     B.~C. Able, R.~A. Tagg, and M.~Rush (1954).
%     Enzyme-catalyzed cellular transanimations.
%     In A.~F.~Round, editor, 
%     {\it Advances in Enzymology Vol. 2} (125--247). 
%     New York, Academic Press.
%  \bibitem[R.~Keohane, 1958]{Keo:58}
%     R.~Keohane (1958).
%     {\it Power and Interdependence: 
%     World Politics in Transition.}
%     Boston, Little, Brown \& Co.
%  \bibitem[Powers, 1985]{Pow:85}
%     T.~Powers (1985).
%     Is there a way out?
%     {\it Harpers, June 1985}, 35--47.

%\end{thebibliography}

%\appendix
%\section{A summary of Latin grammar}    % Each appendix must have a short title.
%\section{Some Latin vocabulary}         % Sections and subsections are supported  
                                        % in the appendices.
\end{document}